\documentclass[aps, prb, reprint, superscriptaddress]{revtex4-1}

\usepackage{amsmath}
\usepackage{graphicx}
\usepackage{graphicx}
\usepackage{amsmath}
\usepackage{amssymb}
\usepackage{url}
\usepackage{bm}		
\usepackage{color}	
\usepackage{textcomp}
\usepackage{xcolor}

\newcommand{\Lcal}{\mathcal{L}}

\newcommand{\rb}{\mathbf{r}}
\newcommand{\xb}{\mathbf{x}}
\newcommand{\yb}{\mathbf{y}}
\newcommand{\vb}{\mathbf{v}}
\newcommand{\cb}{\mathbf{c}}
\newcommand{\drm}{\mathrm{d}}
\newcommand{\Eb}{\mathbf{E}}
\newcommand{\eb}{\mathbf{e}}
\newcommand{\fb}{\mathbf{f}}
\newcommand{\db}{\mathbf{d}}
\newcommand{\Hb}{\mathbf{H}}
\newcommand{\Gb}{\mathbf{G}}
\newcommand{\rhob}{\bm{\rho}}
\newcommand{\epsilonb}{\bm{\epsilon}}
\newcommand{\gb}{\mathbf{g}}
\newcommand{\kb}{\mathbf{k}}
\renewcommand{\oc}[1]{\frac{#1}{c}}
\newcommand{\oct}[1]{\frac{#1}{c^2}}
\newcommand{\eg}{\hat{e}_\mathbf{g}}
\newcommand{\ek}{\hat{e}_\mathbf{k}}
\newcommand{\zh}{\hat{z}}
\newcommand{\gh}{\hat{g}}
\newcommand{\epsb}{\bar{\epsilon}}
\newcommand{\einv}{\frac{1}{\epsilon(\rb)}}
\newcommand{\Th}{\hat{\Theta}}
\newcommand{\Hcal}{\mathcal{H}}
\newcommand{\der}[2]{\frac{\drm #1}{\drm #2}}
\newcommand{\pder}[2]{\frac{\partial #1}{\partial #2}}

\begin{document}

\title{Inverse design of photonic crystals through automatic differentiation}

\author{Momchil~Minkov}
\email[]{mminkov@stanford.edu}

\author{Ian~A.~D.~Williamson}
\affiliation{Department of Electrical Engineering, and Ginzton Laboratory, Stanford University, Stanford, CA 94305, USA}
  
\author{Lucio~C.~Andreani}  
\author{Dario Gerace}
\affiliation{Dipartimento di Fisica, Universit\`a di Pavia, via Bassi 6, 27100 Pavia, Italy}

\author{Beicheng~Lou}
\affiliation{Department of Applied Physics, and Ginzton Laboratory, Stanford University, Stanford, CA 94305, USA}

\author{Alex~Y.~Song}
\affiliation{Department of Electrical Engineering, and Ginzton Laboratory, Stanford University, Stanford, CA 94305, USA}

\author{Tyler~W.~Hughes}
\affiliation{Department of Applied Physics, and Ginzton Laboratory, Stanford University, Stanford, CA 94305, USA}

\author{Shanhui~Fan}
\email[]{shanhui@stanford.edu}
\affiliation{Department of Electrical Engineering, and Ginzton Laboratory, Stanford University, Stanford, CA 94305, USA}

\date{\today}

\begin{abstract}
Gradient-based inverse design in photonics has already achieved remarkable results in designing small-footprint, high-performance optical devices. The adjoint variable method, which allows for the efficient computation of gradients, has played a major role in this success. However, gradient-based optimization has not yet been applied to the mode-expansion methods that are the most common approach to studying periodic optical structures like photonic crystals. This is because, in such simulations, the adjoint variable method cannot be defined as explicitly as in standard finite-difference or finite-element time- or frequency-domain methods. Here, we overcome this through the use of automatic differentiation, which is a generalization of the adjoint variable method to arbitrary computational graphs. We implement the plane-wave expansion and the guided-mode expansion methods using an automatic differentiation library, and show that the gradient of any simulation output can be computed efficiently and in parallel with respect to all input parameters. We then use this implementation to optimize the dispersion of a photonic crystal waveguide, and the quality factor of an ultra-small cavity in a lithium niobate slab. This extends photonic inverse design to a whole new class of simulations, and more broadly highlights the importance that automatic differentiation could play in the future for tracking and optimizing complicated physical models.
\end{abstract}

\maketitle

\section{Introduction}

Tremendous flexibility in the control of the flow of light can be achieved by exploiting the vast number of degrees of freedom in photonic devices with wavelength-scale (or smaller) feature sizes. This flexibility promises the realization of compact and highly efficienct integrated devices, which is becoming increasingly important for future photonic and optoelectronic technologies. 
To take advantage of the degrees of freedom in photonic devices, the field of photonic inverse design has emerged \cite{Molesky2018}, in which an optimization algorithm is used to automate the photonic design process towards a specified device performance as characterized by an objective function. 
This has led to demonstrations of compact devices for routing, wavelength multiplexing, and spatial mode converters \cite{Wang2012, Piggott2015, Frellsen2016, Lin2018}.
Additionally, inverse design has been successfully extended to a number of nonlinear optical devices \cite{Lin2016, Lin2016a, Hughes2018}. 

Gradient-based optimization is probably the most widely used technique in photonic inverse design. In such a technique, within each iteration, one first computes the gradient of the objective function with respect to all the tunable parameters of the device.  
One then varies the parameters along the gradient direction to improve the performance of the device.
Underpinning gradient-based inverse design is the adjoint variable method (AVM), which allows the gradient of a scalar objective function to be efficiently computed with respect to many parameters of the device \cite{Veronis2004, Jensen2011, Lalau-Keraly2013}.
The adjoint variable method can be straightforwardly implemented when the optical devices are simulated through solving a linear system of equations of the form $\hat{A}\eb = \mathbf{b}$, where $\hat{A}$ is the system matrix, $\eb$ is the electromagnetic field distribution in the device to be solved for, and $\mathbf{b}$ is the excitation source. 
In such a case, the gradient  has a very straightforward physical interpretation as the interference between \textit{forward} and \textit{backward} fields.
Such a physical interpretation is useful conceptually, but more importantly, allows for the numerical implementation to use the same solver for both the \textit{forward} and \textit{backward} simulations.
Thus, once a simulation code is set up for the forward simulation, very little additional effort is required to implement the backward simulation.
The mathematical form of the \textit{backward} simulation for computing the gradients is therefore typically derived analytically and hard coded.

There are, however, scenarios in optical device simulations where the dependence of the objective function on the device parameters is more complex. For example, in the design of photonic crystals one often solves an eigenvalue problem $\hat{A} \eb  = \lambda \eb$, where $\lambda$ is an eigenvalue and $\eb$ is the field of the corresponding eigenstate. 
The objective function is then expressed in terms of a mathematical function of some of the eigenvalues and eigenmodes. 
Furthermore, in a mode-expansion method, the matrix $\hat{A}$ itself will typically have a non-trivial dependence on the structural parameters, which has to be accounted for in the gradient.
There are also other examples of simulation techniques such as rigorous coupled wave analysis (RCWA), where the transmission and reflection properties of a multi-layer device are obtained from the eigenstates and eigenvalues in each layer. 
In all these cases, the mathematical dependency of the objective function on the device parameters becomes far more complex. 
Explicit definition and implementation of the backward simulation thus becomes difficult or even infeasible. 
There is therefore a need to systematically implement the adjoint variable method for arbitrary dependency of the objective function on the device parameters. 

In the context of computational science, automatic differentiation (AD) is the application of the adjoint variable method to arbitrary computational graphs.
Within an AD-enabled programming framework, a software developer only needs to define the \textit{forward} computation, while the \textit{backward} computation is generated automatically by tracing program execution or ahead of time via source code analysis.
At the heart of an AD framework are gradient-aware elementary functions which, in essence, can each be represented in terms of their own individual adjoint variable problem, much like the explicitly defined \textit{backward} simulations used in optical inverse design.
However, the key advantage of an AD framework lies in its ability to flexibly compose such elementary functions to build far more complex computations with end-to-end gradient support.

Over the past several decades, automatic differentiation has been explored across a number of contexts \cite{Rumelhart1986, adifor_1992, sambridge_automatic_2007, enciu_automatic_2010, Baydin2018, rackauckas_comparison_2018}.
However, in recent years the growing interest in machine learning, and particularly gradient-based model training, has driven the development modern automatic differentiation libraries.
Several examples include \texttt{Autograd} \cite{autograd}, \texttt{PyTorch} \cite{paszke_pytorch}, \texttt{TensorFlow} \cite{agrawal_tensorflow_2019}, \texttt{Zygote} \cite{innes_don_2019}, and \texttt{JAX} \cite{jax2018github}.
While these libraries have been primarily applied to machine learning, their application to problems in computational physics has only recently been explored \cite{richardson_seismic_2018, hoyer_neural_2019, rackauckas_universal_2020, Hughes2019, Hughes2019a}.
In the context of optical inverse design, an AD-enabled finite difference frequency domain (FDFD) simulation framework was recently proposed \cite{Hughes2019}, which leveraged AD for flexible composition of optimization objective functions and device parameterizations.
Similar goals have motivated the development of application-specific differentiable graph frameworks for optical inverse design \cite{Su2019}.
However, an unexplored application of modern AD frameworks within optical inverse design is to enable gradient computations through optical simulations that do not have straightforwardly defined \textit{backward} problems.
An example class of such simulations are mode expansion methods used for computing the photonic bands of periodic optical structures.

In this work, we present a differentiable implementation of the two-dimensional (2D) plane-wave expansion (PWE) method used for simulating a 2D photonic crystal (PhC), as well as of the guided-mode expansion (GME) method used for efficiently simulating photonic crystal slabs \cite{Andreani2006}. We demonstrate that AD allows us to efficiently compute the gradient of a scalar-valued objective function derived from any output quantities (e.g. eigenmode dispersion, field profile, and/or loss rates) simultaneously with respect to all the input parameters (e.g. position and size of holes, slab permittivity and/or thickness). Our approach is equivalent to an adjoint variable method for the eigenmodes of periodic structures, without the explicit definition of  adjoint fields. Our implementation uses the open source AD package \texttt{Autograd} \cite{autograd}, which allows for great flexibility both in the parametrization of the periodic structures and in the final objective function. Our open-source code has been made available online \cite{legume}. We also show two examples of gradient-based optimizations performed using this method. In the first example, we optimize the dispersion of a PhC waveguide toward several target curves, which is an important problem for nonlinear optics \cite{Boyd2003} and slow light \cite{Baba2008} applications. In the second example, we optimize the quality factor ($Q$) of a PhC cavity in a lithium niobate slab with an ultra-small volume, which can be useful for a number of applications in integrated photonics \cite{Arizmendi2004}. 

\section{Theoretical preliminaries}
\label{sec:theory}

The results presented in this paper are at the intersection of two concepts: automatic differentiation and mode-expansion methods for electromagnetic simulations. In this section, we lay down a theoretical foundation for both of those concepts. We note that throughout this paper, we denote column (``contravariant'') vectors in bold as $\mathbf{v}$, and 2D arrays (matrices) as $\hat{M}$. Derivatives with respect to a vector ($\drm/\drm \vb)$ are therefore row (``covariant'') vectors.

\subsection{Automatic differentiation}
\label{sec:ad}

\begin{figure}
\centering
\includegraphics[width=0.48\textwidth]{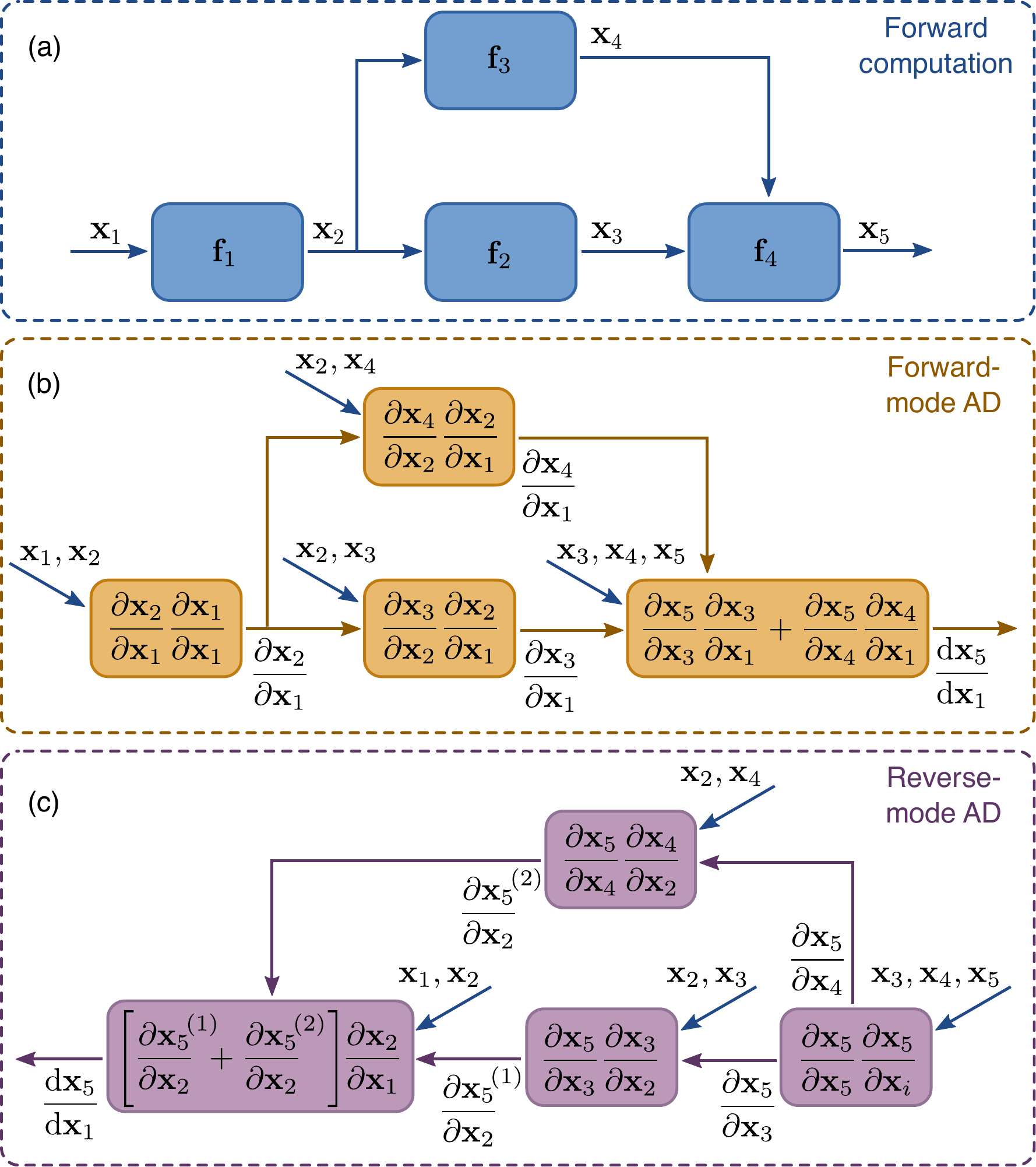}
\caption{(a): An example computer program represented as a computational graph. (b): Forward-mode differentiation graph associated to (a). (c): Revere-mode differentiation graph associated to (a). In (b) and (c), blue arrows denote inputs from the forward computation of (a).}
\label{fig:autodiff}
\end{figure}

The field of automatic differentiation has existed for more than five decades \cite{Wengert1964, Griewank2008, Baydin2018} and has covered a wide range of applications. Our goal here is only to provide a summary targeted at readers with little or no prior knowledge of the topic, and hence many details are omitted for the sake of brevity. The first important point is that AD is a computer programming paradigm, and not a purely mathematical construct. That is to say, unlike symbolic differentiation, automatic differentiation is inextricably connected to an underlying computer program, which can be generically represented as a computational graph, as shown in Fig. \ref{fig:autodiff}(a). In the schematic, the rectangles denote generic functions included in a programming library, while vectors $\mathbf{x}_i$ contain all the inputs and outputs of these functions. Thus, the \textit{forward computation} of the program illustrated in Fig. \ref{fig:autodiff}(a) performs the operations
\begin{align*}
    &\xb_2 = \fb_1(\xb_1) \\
    &\xb_3 = \fb_2(\xb_2) \\
    &\xb_4 = \fb_3(\xb_2) \\
    &\xb_5 = \fb_4(\xb_3, \xb_4).
\end{align*}

An automatic differentiation library then allows the user to compute the \textit{exact} Jacobian $\mathrm{d}\mathbf{x}_i/\mathrm{d}\mathbf{x}_j$ for any $i, j$ using the rules of differentiation and some knowledge of the partial derivatives of each operation (more on this below). For example, in the case of the computational graph of Fig. \ref{fig:autodiff}(a), we have
\begin{equation}
    \der{\xb_5}{\xb_1} = \left[\pder{\xb_5}{\xb_4}\pder{\xb_4}{\xb_2}  + \pder{\xb_5}{\xb_3}\pder{\xb_3}{\xb_2} \right]\pder{\xb_2}{\xb_1}, 
    \label{eqn:ad_example}
\end{equation}
which is in itself a computational graph that can be traced by the program. A crucial point in AD is that this tracing depends on the order in which the products in eq. (\ref{eqn:ad_example}) are computed (the ``accumulation'' of the Jacobian), which also affects the computational complexity of evaluating eq. (\ref{eqn:ad_example}). There are two main approaches. The first one is to work from right to left, which is called forward-mode (FM), because the Jacobian accumulation, illustrated in Fig. \ref{fig:autodiff}(b), follows the arrows of the original computational graph of Fig. \ref{fig:autodiff}(a). This is conceptually straightforward, as, for example, fan-out of the outputs (e.g. $\xb_2$ in Fig. \ref{fig:autodiff}(a)) leads to fan-out of the derivative, while fan-in of the inputs (e.g. $\xb_3$ and $\xb_4$ in Fig. \ref{fig:autodiff}(a)) leads to addition of derivatives. The alternative approach is to work from left to right in eq. (\ref{eqn:ad_example}), which is called reverse-mode (RM) since the computational graph (Fig. \ref{fig:autodiff}(c)) flows in the opposite direction of the arrows in Fig. \ref{fig:autodiff}(a). In this case, everything is reversed, and fan-out of the output leads to addition of the input derivatives ($\partial \xb_5^{(1)}/\partial \xb_2$ and $\partial \xb_5^{(2)}/\partial \xb_2$ in Fig. \ref{fig:autodiff}(c)), while fan-in of the inputs leads to splitting of the output derivatives ($\partial \xb_5/\partial \xb_4$ and $\partial \xb_5/\partial \xb_3$ in Fig. \ref{fig:autodiff}(c)). More generally, a combination of FM and RM is possible, but this is not typically used due to the extra complexity. 

The construction of the computational graph for evaluating the derivatives is one of the main components of an AD library, and there are various ways in which it can be implemented. The specifics of this lie beyond the scope of our discussion, but the important point is that once the graph is constructed, the derivative computation in both forward- and reverse-mode accumulation becomes a sequence of elementary building-block operations, which we discuss in more detail below.

The forward-mode derivative computation corresponding to Fig. \ref{fig:autodiff}(a) is shown in Fig. \ref{fig:autodiff}(b). There, brown arrows indicate gradient flow, i.e. input and output of partial derivatives $\partial \xb_i / \partial \xb_j$, while blue arrows indicate inputs from the forward computation, i.e. $\xb_i$. We see that the primitive building block of the forward-mode computation in Fig. \ref{fig:autodiff}(b) is the operation
\begin{equation}
g_{FM}\left(\xb_i, \xb_j, \pder{\xb_j}{\xb_1}\right) = \pder{\xb_i}{\xb_j} \pder{\xb_j}{\xb_1}.
\label{eqn:gfm}
\end{equation}
Notice that in eq. (\ref{eqn:gfm}), the right-most differentiation always has $\xb_1$ in the denominator. Then, computing $\drm \xb_5 / \drm \xb_1$ in our example requires $N_1$ separate evaluations of the graph of Fig. \ref{fig:autodiff}(b), where $N_1$ is the number of elements in $\xb_1$. As a general rule, the computational time of forward-mode AD scales linearly with the number of \textit{input} parameters. In many implementations (including \texttt{Autograd}), this is made explicit by considering the forward accumulation of a single derivative vector $\partial \xb_j/\partial x$, and associating a Jacobian-vector product (JVP) function to every function $\fb(\xb)$, such that
\begin{equation}
\mathrm{jvp}(\xb, \fb(\xb), \vb) = \pder{\fb}{\xb} \vb.
\label{eqn:jvp}
\end{equation}
The vector $\vb$ in eq. (\ref{eqn:jvp}) is essentially a single column of the Jacobian $\partial \xb_j / \partial \xb_1$ that appears in eq. (\ref{eqn:gfm}). Given a set of functions with correspondingly defined $\mathrm{jvp}$-s, arbitrarily complex programs can be differentiated using forward-mode AD. 

The reverse-mode derivative accumulation for the same computation is shown in Fig. \ref{fig:autodiff}(c). We see that the primitive building block is the operation
\begin{equation}
g_{RM}\left(\xb_i, \xb_j, \pder{\xb_5}{\xb_i}\right) = \pder{\xb_5}{\xb_i}\pder{\xb_i}{\xb_j} .
\label{eqn:grm}
\end{equation}
Here, the left-most differentiation always has $\xb_5$ in the numerator. Then, computing $\drm \xb_5 / \drm \xb_1$ in our example requires $N_5$ separate evaluations of the graph of Fig. \ref{fig:autodiff}(c), and, as a general rule, the computational time of reverse-mode AD scales linearly with the number of \textit{outputs} of the program. One possible implementation of eq. (\ref{eqn:grm}) is to associate a vector-Jacobian product (VJP) to every function $\fb(\xb)$, such that
\begin{equation}
\mathrm{vjp}(\xb, \fb(\xb), \vb) =  \vb^T \pder{\fb}{\xb}.
\end{equation}

There are two crucial differences between forward- and reverse-mode AD. First, regardless of the implementation, the computational time for $\drm \xb_o / \drm \xb_i$ scales with $N_o$ when using reverse-mode and with $N_i$ when using forward-mode. Here, the ``output'' vector $\xb_o$ and the ``input'' vector $\xb_i$ can be any of the variables along the computational graph, i.e. any one of $\xb_1, \xb_2, \dots \xb_5$ in the example of Fig. \ref{fig:autodiff}(a). Thus, when $N_o \gg N_i$, FM is much faster than RM, and vice-versa if $N_i \gg N_o$. Since optimization problems typically have a large number of input parameters and a single scalar-valued objective function as output, reverse-mode AD is then substantially better for these problems. This explains the ubiquitous use of backpropagation in machine learning and of the adjoint variable method in engineering (both are specific examples of reverse-mode AD). The second difference, however, works against reverse-mode AD, and can sometimes be a limiting factor. Specifically, note that all intermediate results from the forward computation (blue arrows in Fig. \ref{fig:autodiff}(b)-(c)) are needed for the derivative computation in both FM and RM AD. Since the forward-mode AD can be done in parallel with the forward computation, these do not need to be stored beyond every individual step. In contrast, the reverse-mode computation can only start after the forward computation is complete, and requires that all intermediate results are stored. This could lead to a significantly higher memory requirement, which has motivated the development of checkpointing schemes \cite{Dauvergne2006}, however a detailed discussion of such schemes is beyond the scope of this paper. That said, the computational time advantage of reverse-mode AD is so compelling that the typical approach is to accommodate its memory disadvantage in some way.

The \texttt{Autograd} library is API-compatible with a large subset of the \texttt{NumPy} and \texttt{SciPy} libraries \cite{numpy_2011,virtanen_scipy_2020}. Thus, complex programs that include algebraic operations, array manipulations, and flow control can be traced and automatically differentiated with ease. The end user needs only to write the forward computation (as in Fig. \ref{fig:autodiff}(a)), and the differentiation is handled automatically. This gives enormous flexibility in defining objective functions and parametrizations in numerical simulations and optimizations. In addition, external functions can also be included if the JVP and/or VJP can be defined. We will illustrate this with a specific example of a function that will be needed for the guided-mode expansion of Section \ref{sec:gme}. Furthermore, this same example captures the essence of the adjoint variable method for finite-difference frequency-domain electromagnetic simulations for both linear and nonlinear systems \cite{Veronis2004, Jensen2011, Lalau-Keraly2013, Hughes2019}.

Assume that the output $\xb_o$ of an operation is defined through a set of constraints written most generally as
\begin{equation}
    \fb(\xb_i, \xb_o) = 0.
    \label{eqn:constraint}
\end{equation}
We assume $\fb$ has the same dimensionality as $\xb_o$ and that a solution (not necessarily unique) exists, and can be found through some iterative numerical method. Then, $\xb_o$ is an implicit function of the inputs $\xb_i$, and using the differentiation rule for implicit functions we get
\begin{equation}
    \pder{\xb_o}{\xb_i} = -\left[\pder{\fb}{\xb_o}\right]^{-1} \pder{\fb}{\xb_i},
    \label{eqn:implicit_der}
\end{equation}
where $\left[\partial \fb/\partial \xb_o \right]^{-1}$ is the matrix inverse of the square Jacobian matrix $\partial \fb/\partial \xb_o$. Since we are mostly interested in reverse-mode AD, we will illustrate the VJP computation for this function, which is
\begin{equation}
    \vb_o = \vb_i^T \pder{\xb_o}{\xb_i} = -\vb_i^T \left[\pder{\fb}{\xb_o}\right]^{-1} \pder{\fb}{\xb_i}.
\end{equation}
Thus, the output $\vb_o$ of the VJP  can be found using the auxiliary variable $\vb_a$ that is found as the solution to the \textit{linear} system of equations
\begin{equation}
    \left[\pder{\fb}{\xb_o}\right]^T \vb_a = -\vb_i, 
    \label{eqn:vjp_fsolve}
\end{equation}
and reads
\begin{equation}
    \vb_o = \vb_a^T \pder{\fb}{\xb_i}.
    \label{eqn:fsolve_adj}
\end{equation}
Note that $\vb_o$ has the same dimension as $\xb_i$, while $\vb_i$ and $\vb_a$ have the same dimension as $\xb_o$, since input and output dimensions are switched in reverse-mode AD.

The essence of the adjoint variable method for an electromagnetic finite-difference frequency domain simulation is captured by eqs. (\ref{eqn:vjp_fsolve}-\ref{eqn:fsolve_adj}). In such a simulation, we have a discretized real-space domain and an input vector $\xb_i = \epsilonb$ defining the dielectric permittivity at every point. The output is the corresponding electric field $\xb_o = \eb$ at every point, which is obtained by solving for $\fb(\epsilonb, \eb) = 0$, where $\fb$ defines the Maxwell problem. For a linear electromagnetic system, $\fb(\epsilonb, \eb) = \hat{A}(\epsilonb) \eb - \mathbf{b}$, with $\hat{A}(\epsilonb)$ the matrix defining the single-frequency linear Maxwell's equations on a discretized spatial grid, and $\mathbf{b}$ a current source term \cite{Veronis2004, Jensen2011, Lalau-Keraly2013}. In the case of e.g. Kerr-type nonlinearity, $\fb(\epsilonb, \eb) = \hat{A}(\epsilonb, \eb) \eb - \mathbf{b}$ and a nonlinear numerical solver is required \cite{Hughes2019}. However, the VJP in both cases works in exactly the same way, as defined in eq. (\ref{eqn:fsolve_adj}). The auxiliary variable $\vb_a$ is called the adjoint field in the adjoint variable method terminology. It is actually worth noting that any mathematical operation can be framed as a constraint -- e.g. $\xb_o = \xb_i^2$ is equivalent to $\fb(\xb_i, \xb_o) = \xb_i^2 - \xb_0 = 0$. Therefore, any VJP can be derived through the implicit function theorem and eqs. (\ref{eqn:constraint}-\ref{eqn:fsolve_adj}). In that sense, reverse-mode AD is sometimes derived within the framework of Lagrange multipliers, which end up corresponding to the adjoint variables \cite{Griewank2008}.

In conclusion, we wish to point out that automatic differentiation, and specifically reverse-mode AD, is more than just a numerical subtlety with limited importance. On the contrary, when considering optimization problems, the method is computationally faster than \textit{both} numerical differentiation (i.e. finite-difference methods) and symbolic differentiation (deriving by hand, and evaluating the final expression for the derivative). The improvement is due to the use of intermediate ``adjoint'' variables. The second important advantage of AD is splitting operations into elementary building blocks, which allows arbitrarily complex programs to be differentiated. For these reasons, the development and use of backpropagation, which corresponds to reverse-mode AD, has been the key driver of the recent machine learning revolution \cite{Baydin2018}, and we believe that it will be a key component of tackling optimization problems in physics and engineering.

\subsection{Mode expansion methods}
\label{sec:me}

In this section, we review the general idea of mode expansion methods applied to Maxwell's equations. The plane-wave expansion (Section \ref{sec:pwe}) and the guided-mode expansion (Section \ref{sec:gme}) are both specific examples of the formalism we introduce below.

In the absence of free charges and currents, Maxwell's equations can be written as an eigenvalue problem for the electric ($\Eb$) and magnetic ($\Hb$) fields with harmonic time-dependence, $\Eb(\rb, t) = \Eb(\rb) e^{-i\omega t}$, $\Hb(\rb, t) = \Hb(\rb) e^{-i\omega t}$ \cite{Joannopoulos2008}. If we further assume a linear, isotropic, lossless, non-dispersive and non-magnetic medium, which is a good approximation for many real-world materials, the eigenvalue equation can be written for $\Hb$ alone:
\begin{equation}
    \Th \Hb(\rb) \equiv \left[ \nabla \times \einv \nabla \times \right] \Hb(\rb) = \oct{\omega^2} \Hb(\rb),
    \label{eqn:max_eig}
\end{equation}
together with the constraint $\nabla \cdot \Hb (\rb) = 0$. In eq. (\ref{eqn:max_eig}), $\epsilon(\rb)$ is the relative dielectric permittivity distribution that fully defines the system. For a known magnetic field profile, the electric field can be found through
\begin{equation}
    \Eb(\rb) = \frac{ic}{\omega \epsilon(\rb)} \nabla \times \Hb(\rb). \label{eqn:EfromH}   
\end{equation}

For a closed system in which all fields decay as $|\rb| \rightarrow \infty$, or for a periodic system, we can define an inner product on the space of magnetic field functionals as
\begin{equation}
    (\Hb_\nu, \Hb_\mu) = \int \drm \rb \Hb_\nu^\dagger(\rb) \Hb_\mu(\rb).
    \label{eqn:inner_prod}
\end{equation}
such that the operator $\Th \Hb(\rb)$ is Hermitian, i.e. $(\Th\Hb_\nu, \Hb_\mu)^* = (\Hb_\nu, \Th\Hb_\mu)$. For an open system, the integration of eq. (\ref{eqn:inner_prod}) is not well defined, and correctly defining an inner product is an active area of study that goes beyond the scope of this work \cite{Lalanne2018}. Here, we will assume either periodic or decaying fields, and compute concrete examples of eq. (\ref{eqn:inner_prod}) in Sections \ref{sec:pwe} and \ref{sec:gme}. 

The idea of a mode expansion method is to express the eigenstates of eq. (\ref{eqn:max_eig}) for an arbitrary operator $\Th$ in the basis of the eigenstates of a different operator $\Th_0$, i.e.
\begin{equation}
    \Hb(\rb) = \sum_\mu c_\mu \Hb_\mu(\rb),
    \label{eqn:mode_exp}
\end{equation}
where $\Th_0\Hb_\mu = (\omega_\mu^2/c^2)\Hb_\mu$ and $\nabla \cdot \Hb_\mu = 0$ holds for all $\mu$. In other words, the modes $\Hb_\mu$ are solutions to Maxwell's equations for a given structure which is typically simple to solve, e.g. $\epsilonb(\rb) =  1$ everywhere (free space). Because the problem is Hermitian, the eigenmodes form an orthonormal set, such that
\begin{equation}
    (\Hb_\nu, \Hb_\mu) = \delta_{\nu\mu}.
\end{equation}
Using this, plugging eq. (\ref{eqn:mode_exp}) into eq. (\ref{eqn:max_eig}), multiplying by $\Hb_\nu^*$ and taking the inner product on both sides, we get
\begin{equation}
    \left(\Hb_\nu, \Th \sum_\mu c_\mu \Hb_\mu\right) = \sum_\mu c_\mu \oct{\omega^2} (\Hb_\nu, \Hb_\mu)
\end{equation}
and so
\begin{equation}
    \sum_\mu \Hcal_{\nu\mu} c_\mu = \oct{\omega^2} c_\nu.
    \label{eqn:eigenproblem}
\end{equation}
This is an eigenproblem for the expansion coefficients $c_\mu$ with matrix elements $\mathcal{H}_{\nu\mu} = (\Hb_\nu, \Th \Hb_\mu)$. 

We note that the expansion of eq. (\ref{eqn:mode_exp}) is exact only when $\Hb_\mu$ forms a complete set, and all modes are included in the summation. However, any starting structure $\Th_0$ has an infinite number of eigenmodes, with increasing magnitude of the eigenfrequency. In practice, the summation will thus always be truncated to a finite subset of the basis functions. Apart from that, the set of plane waves used in Section \ref{sec:pwe} is complete, and can be thought of as Fourier decomposition of a vector-valued function. On the other hand, in the guided-mode expansion method of Section \ref{sec:gme}, we restrict the summation to the fully-guided slab modes, which is an \textit{incomplete} set. The method is thus only approximate, but it has been shown to compare well to first-principle simulations for a number of different structures, with a much faster computational speed \cite{Minkov2014, Minkov2015, Minkov2018}.

\section{Plane-wave expansion}
\label{sec:pwe}

The plane-wave expansion method for simulating periodic structures is well-known both in photonics and in quantum mechanics and other domains. Here, we will nevertheless review the fundamentals, as this helps with the understanding of the guided-mode expansion of Section \ref{sec:gme}, as well as the understanding of our automatic differentiation implementation of the two methods. 

\subsection{Method description}

\begin{figure}
\centering
\includegraphics[width=0.48\textwidth]{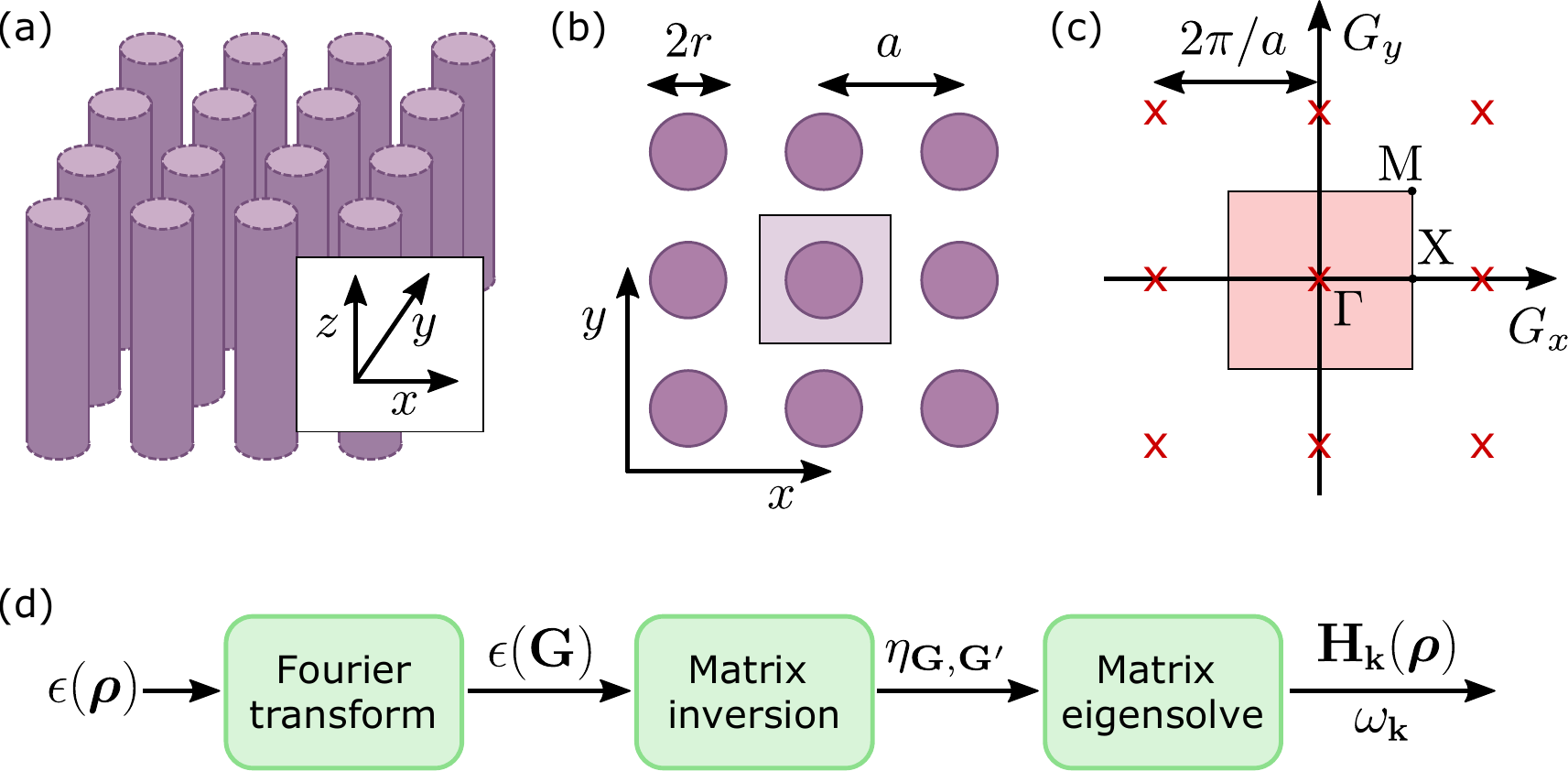}
\caption{(a): A 2D photonic crystal composed of dielectric rods in air. The rods form a square lattice in the $xy$-plane, and are assumed to be infinite in the $z$-direction. (b): Cross-section in the $xy$-plane showing the lattice constant $a$ and rod radius $r$. The shaded square shows the primitive cell of the crystal. (c): Reciprocal lattice for this crystal, with reciprocal lattice vectors separated by $2\pi/a$ in both directions. The shaded square shows the primitive cell, i.e. the first Brillouin zone, and the high-symmetry points $\Gamma$, X and M are noted. (d): Computational graph for the 2D plane-wave expansion.}
\label{fig:pwe}
\end{figure}

In this paper, we focus on 2D PWE, in which the simulation domain is assumed to be periodic in two directions and translationally invariant in the third. One example of such a structure is illustrated in Fig. \ref{fig:pwe}(a)-(b) and consists of an array of dielectric rods in air, assumed to be infinitely extended in the $z$-direction. Most generally, the Bloch theorem then states that the eigenmodes of the structure can be written as 
\begin{equation}
    \Hb_\kb(\rhob) = e^{i k_z z} e^{i\kb \cdot \rhob} \tilde{\Hb}_{\kb, k_z}(\rhob),
\end{equation}
where the Bloch momentum $\kb$ and the position vector $\rhob$ lie in the $xy$-plane, and $\tilde{\Hb}_{\kb, k_z}(\rhob)$ is periodic in $xy$ with the 2D lattice periodicity. As is usual in the study of 2D PhCs, we will focus on modes with $k_z = 0$ and drop that label. The method is easy to generalize for non-zero $k_z$.

The inner product (\ref{eqn:inner_prod}) for this 2D problem reads
\begin{equation}
    (\Hb_\nu, \Hb_\mu) = \frac{1}{S}\int_S \drm \rhob \Hb_\nu^\dagger(\rhob) \Hb_\mu(\rhob),
    \label{eqn:pwe_prod}
\end{equation}
where $S$ denotes the primitive cell in the $xy$-plane. With this, we can apply the general mode-expansion procedure outlined in section \ref{sec:me}. Because of the reflection symmetry with respect to the $xy$-plane, all modes can be separated into transverse electric (TE) and transverse magnetic (TM) polarizations, even in the case of arbitrary in-plane permittivity $\epsilon(\rhob)$. The plane-wave basis is just the solution to the free-space system with relative permittivity $\epsilon(\rhob) = 1$ everywhere. This can be written as $\Hb_\kb(\rho) = \Hb^p_\kb e^{i\kb \cdot \rhob}$, where $\Hb^p_\kb = \zh$ for TE polarization, while $\Hb^p_\kb = \ek$ for TM polarization. Here, $\zh$ and $\ek$ are unit vectors pointing respectively in the $z$-direction and in the direction orthogonal to both $z$ and $\kb$. The $p$-polarization eigenstates for an arbitrary 2D PhC can thus be written as
\begin{equation}
    \Hb_\kb^p(\rhob) = \sum_\Gb c_\kb^p(\Gb) \Hb^p_{\Gb + \kb} e^{i(\Gb + \kb) \cdot \rhob},
    \label{eqn:pwe_exp}
\end{equation}
where $\Gb$ is a reciprocal lattice vector as shown in Fig. \ref{fig:pwe}(c). Using the inner product of eq. (\ref{eqn:pwe_prod}) and the definition of the Maxwell operator in eq. (\ref{eqn:max_eig}), the matrix elements as defined in eq. (\ref{eqn:eigenproblem}) for the two polarizations read
\begin{align}
    \Hcal_{\Gb, \Gb'}^{\mathrm{TE}} &= [(\kb + \Gb)\cdot (\kb + \Gb')] \eta_{\Gb,\Gb'}, \label{eqn:pwe_te}\\
    \Hcal_{\Gb, \Gb'}^{\mathrm{TM}} &= |\kb + \Gb| |\kb + \Gb'| \eta_{\Gb,\Gb'}, \label{eqn:pwe_tm}
\end{align}
where $|\vb|$ is the magnitude of vector $\vb$. Furthermore, we denote the 2D Fourier transform of the inverse of the permittivity distribution by the matrix elements
\begin{equation}
    \eta_{\Gb, \Gb'} = \frac{1}{S} \int_S \drm \rhob \frac{1}{\epsilon(\rhob)} e^{- i (\Gb-\Gb') \cdot \rhob}.
    \label{eqn:eps_ft}
\end{equation}
In the expansion of eq. (\ref{eqn:pwe_exp}), we only include a discrete set of plane waves defined by the reciprocal lattice vectors, because the Fourier components in eq. (\ref{eqn:eps_ft}) are zero for any plane-wave combination that is non-commensurate with the lattice periodicity.

Numerically, we truncate the expansion up to some maximum value $G_\mathrm{max}$, such that $|\Gb| \le G_\mathrm{max}$. Furthermore, we note that it has been shown that, due to the discontinuous nature of the permittivity, a significantly better convergence of the expansion is achieved if the Fourier transform $\epsilon(\Gb)$ of $\epsilon(\rhob)$ is computed first, and $\hat{\eta}$ is computed as the matrix inverse of the matrix $\epsilon_{\Gb, \Gb'} = \epsilon(\Gb - \Gb')$ \cite{Ho1990, Li1996}. A high-level computational graph for the full plane-wave expansion is then presented in Fig. \ref{fig:pwe}(d), showing the three main operations needed to compute the eigenmodes of an arbitrary 2D PhC. These are: Fourier transform of the real-space permittivity to compute $\epsilon(\Gb)$; matrix inversion to compute $\eta_{\Gb, \Gb'}$; and matrix diagonalization of $\hat{\Hcal}$ as defined in eq. (\ref{eqn:pwe_te})-(\ref{eqn:pwe_tm}) for the two polarizations. Next, we describe the automatic differentiation aspect of the PWE method.

\subsection{Automatic differentiation}
\label{sec:pwe_ad}

As emphasized in Section \ref{sec:ad}, reverse-mode AD is the method of choice for optimization problems, which is why from here on we only focus on this method. Thus, we discuss the derivative propagation through the operations of Fig. \ref{fig:pwe}(d) in reverse order, starting with the matrix eigensolve. Generically, we denote the forward operation as a function $\mathrm{eigh}(\hat{\Hcal})$ that takes a Hermitian matrix $\hat{\Hcal}$ as an input and returns a set of eigenvalues $E_i$ and associated eigenvectors $\cb_i$ such that 
\begin{equation}
    \hat{\Hcal} \cb_i = E_i \cb_i, \quad \quad \forall i.
    \label{eqn:eigh}
\end{equation}
In the current version (1.3) of \texttt{Autograd}, the $\mathrm{eigh}$ operation is only implemented for real symmetric matrices. We thus extended the library by defining the reverse-mode AD step for the eigensolve of a Hermitian matrix. This operation can be derived using matrix algebra \cite{Giles2008, Lee2007}, or using eqs. (\ref{eqn:constraint}-\ref{eqn:implicit_der}) and treating eq. (\ref{eqn:eigh}) as a constraint. Below, we give yet another, more intuitive derivation using perturbation theory from quantum mechanics.

\begin{figure*}
\centering
\includegraphics[width=\textwidth]{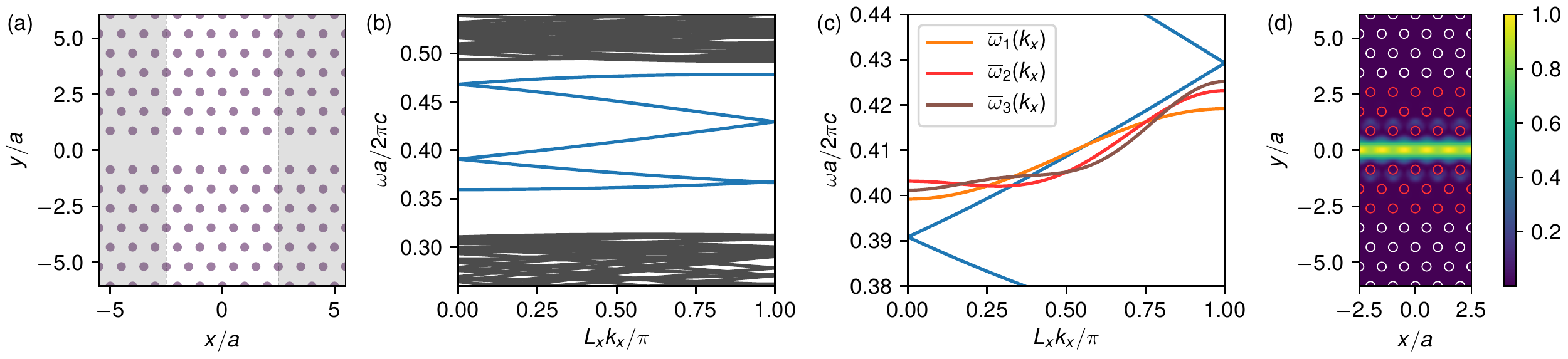}
\caption{(a): Photonic crystal waveguide made by a missing row of pillars in a hexagonal lattice of dielectric pillars of relative permittivity $\epsilon = 9$ and radius $r = 0.2a$, with $a$ the lattice constant. The white region of width $L_x = 5a$ in the $x$-direction denotes the supercell used in the plane-wave expansion simulations. (b): Photonic bands of the structure; waveguide bands are shown in blue. (c): Zoom-in over the middle of the waveguide-band region, and three different target dispersion curves that we optimize for. (d): Electric field intensity for the mode at $L_x k_x/\pi = 0.5$ for the blue band in (c). The positions and radii of the pillars marked in red are optimized to achieve the target dispersion curves in (c).}
\label{fig:wg_start}
\end{figure*}

We assume that there are no degenerate eigenvalues. Degenerate cases can also be handled by perturbation theory, but this requires extra considerations and is neglected for simplicity. Then, the eigenvalues can be labeled (e.g. sorted by magnitude), and derivative of a given eigenvalue with respect to a given matrix element is 
\begin{equation}
    \pder{E_i}{\Hcal_{\mu \nu}} =  \lim_{\lambda \rightarrow 0} \frac{E_i(\hat{\Hcal} + \lambda \hat{1}_{\mu\nu}) - E_n(\hat{\Hcal})}{\lambda},
\end{equation}
where $E_i(\Hcal)$ is the $i$-th eigenvalue of matrix $\hat{\Hcal}$, and $\hat{1}_{\mu\nu}$ is a matrix with the same dimension as $\hat{\Hcal}$ with zeros everywhere apart from the $\mu, \nu$ element, which is set to one. Since we take a limit of $\lambda \rightarrow 0$, we can use first-order perturbation theory as an exact result:
\begin{equation}
    E_i(\hat{\Hcal} + \lambda \hat{1}_{\mu\nu}) = E_i(\hat{\Hcal}) + \lambda \cb_i^\dagger  \hat{1}_{\mu\nu}\cb_i.
\end{equation}
Therefore,
\begin{equation}
    \pder{E_i}{\Hcal_{\mu \nu}} = \lim_{\lambda \rightarrow 0} \frac{\lambda \cb_i^\dagger \hat{1}_{\mu\nu}\cb_i}{\lambda} = c_{i, \mu}^*c_{i, \nu}, 
\end{equation}
where $c_{i, \mu}$ is the $\mu$-th element of $\cb_i$. The VJP for the eigensolve can then be written as a matrix with elements
\begin{equation}
\mathrm{vjp}(\hat{\Hcal}, \mathrm{eigh}(\hat{\Hcal}), \vb)_{\mu \nu} = \sum_i v_i c_{i, \mu}^*c_{i, \nu}.
\end{equation}
We can also use first-order perturbation theory to compute derivatives of the eigenvectors:
\begin{align}
    \pder{\cb_{i}}{\Hcal_{\mu \nu}} &=  \lim_{\lambda \rightarrow 0} \frac{\cb_{i}(\Hcal + \lambda \hat{1}_{\mu\nu}) - \cb_{i}(\Hcal)}{\lambda} \label{eqn:eigvec_der} \\ \nonumber
    &= \lim_{\lambda \rightarrow 0} \sum_{j\neq i} \frac{\cb_j^\dagger \lambda \hat{1}_{\mu\nu}\cb_i}{\lambda(E_i - E_j)} \cb_j = \sum_{j\neq i} \frac{c_{j, \mu}^*c_{i, \nu}}{E_i - E_j} \cb_j, 
\end{align}
and write the corresponding VJP.

Something very important to notice is the fact that the derivative of an eigenvalue depends only on its corresponding eigenvector. On the other hand, the derivative of an eigenvector depends on all other eigenvectors of the matrix, which could have important implications. In particular, if an objective function depends on any of the eigenvectors, \textit{all} of them are needed to propagate the gradient \textit{exactly}. However, computing all eigenvectors is difficult when using iterative methods that return only a few eigenvectors \cite{Johnson2001}. Still, the dependence of the denominator in eq. (\ref{eqn:eigvec_der}) on $E_i - E_j$ means that eigenvectors with very different eigenvalues should contribute less to the gradient. As a general statement, the backpropagation step through an eigensolve, in which only a subset of the eigenmodes are computed, is exact for the eigenvalues, but is only approximate for the eigenvectors (becoming exact in the limit of computing all eigenvectors).

The Hermitian eigenvalue decomposition is the only operation that required extending the existing \texttt{Autograd} API. All other operations, including a number of low-level algebraic operations that are not shown in Fig. \ref{fig:pwe}(d), are already supported by \texttt{Autograd}. For completeness, here we briefly discuss the other two main operations of the computational graph shown in Fig. \ref{fig:pwe}(d). First, the AD step through the matrix inverse can be derived starting from $\drm \hat{A}^{-1}/\drm p = - \hat{A}^{-1} \drm \hat{A} /\drm p \hat{A}^{-1}$, and the VJP reads \cite{Giles2008}
\begin{equation}
    \mathrm{vjp} \left(\hat{A}, \hat{A}^{-1}(\hat{A}), \hat{V} \right) = -\hat{A}^{-1}\hat{V}^T\hat{A}^{-1}, 
\end{equation}
where $\hat{V}$ is now a matrix of the same size as $\hat{A}$. This is a convenient way to write the expression that can be assumed without loss of generality -- strictly speaking, we can always ``flatten'' $\hat{A}$ and $\hat{V}$ into vectors to more rigorously match the formalism as described in Section \ref{sec:ad}. Second, regarding the Fourier transform, one way that this can be implemented in practice is to define $\epsilon(\rhob)$ on a discrete grid in real space, and use a discrete Fourier transformation (DFT). In that case, we simply have $\epsilon(\Gb) = \hat{D}\epsilon(\rhob)$, where $\hat{D}$ is the corresponding DFT matrix, and the VJP through this linear operation is straightforward, $\vb_o^T = \vb_i^T \hat{D}$.

In our implementation \cite{legume}, we take a different approach to the parametrization of the permittivity. Namely, we define the structure through \textit{shape} primitives like circles and polygons. Then, the Fourier transform can be computed through simple algebraic operations \cite{Andreani2006, Lee1983} that are straightforward to differentiate through. For circles, the Fourier transform also includes the Bessel function of the first kind, whose derivative is analytic and the function is already included in \texttt{Autograd}. In Section \ref{sec:pwe_opt} below, we demonstrate in practice this parametrization, and the entire formalism developed thus far.

\subsection{Waveguide optimization}
\label{sec:pwe_opt}

In this section we apply the PWE and AD formalism developed in the previous sections to a practical design problem. Namely, we optimize the dispersion $\omega(\kb)$ of a PhC waveguide to match various pre-defined target forms. Dispersion engineering in general is important for a number of practical applications including nonlinear phase matching \cite{Boyd2003} and generation of frequency combs \cite{Moss2013}. Here, for illustrative purposes, we use generic target dispersion curves. The starting structure is shown in Fig. \ref{fig:wg_start}(a), and consists of a row of missing holes in a hexagonal lattice of dielectric rods in air. The underlying bulk PhC has a broad band gap for TM-polarized modes, and the line defect introduces a guided band inside the band gap, as shown in Fig. \ref{fig:wg_start}(b). Note that the guided band (blue) is folded because we use a supercell of size $L_x = 5a$ in the propagation direction, such that within a supercell there are sufficient degrees of freedom for optimization purposes. 

In Fig. \ref{fig:wg_start}(c), we show three different target dispersion curves for which we will optimize the waveguide. These are defined by a set of increasing Fourier components in $k$-space as
\begin{align}
    \bar{\omega}_1(k_x) &= -0.01\cos(k_x L_x), \\
    \bar{\omega}_2(k_x) &= -0.01\cos(k_x L_x) + 0.004\cos(2k_x L_x), \\
    \bar{\omega}_3(k_x) &= -0.01\cos(k_x L_x) + 0.004\cos(2k_x L_x) \\ \nonumber
    & \quad \quad \quad - 0.002\cos(3k_x L_x),
\end{align}
where we define $\bar{\omega}$ as the dimensionless reduced frequency $\omega a/2\pi c$. We take only cosine components because, due to time-reversal invariance, the dispersion must have zero derivative at the high-symmetry Brillouin zone points. The three target curves are illustrated in Fig. \ref{fig:wg_start}(c), on top of the middle of the five guided bands of the waveguide, which is the one we optimize to match the targets. We allow for an arbitrary offset in frequency, and define the objective function as a mean-square error (MSE): 
\begin{equation}
    \Lcal = \int^{L_x}_0 \drm k_x |\bar{\omega}(k_x) - \langle \bar{\omega} \rangle - \bar{\omega}_i(k_x)|^2,
\end{equation}
where $\bar{\omega}(k_x)$ is the reduced frequency of the waveguide band, and $\langle \bar{\omega} \rangle$ denotes $\bar{\omega}(k_x)$ averaged over $k_x$. As optimization parameters, we take the position and radii of the three rows of rods around the waveguide, as illustrated in Fig. \ref{fig:wg_start}(d). The shifts are applied symmetrically w.r.t. the $y=0$ plane, such that there are 45 free parameters total ($\Delta x$, $\Delta y$, and $\Delta r$ for 15 rods on one side of the waveguide). The electric field intensity $|E_z|^2$ for one of the guided modes of the starting waveguide is also shown in Fig. \ref{fig:wg_start}(d).

\begin{figure}
\centering
\includegraphics[width=0.48\textwidth]{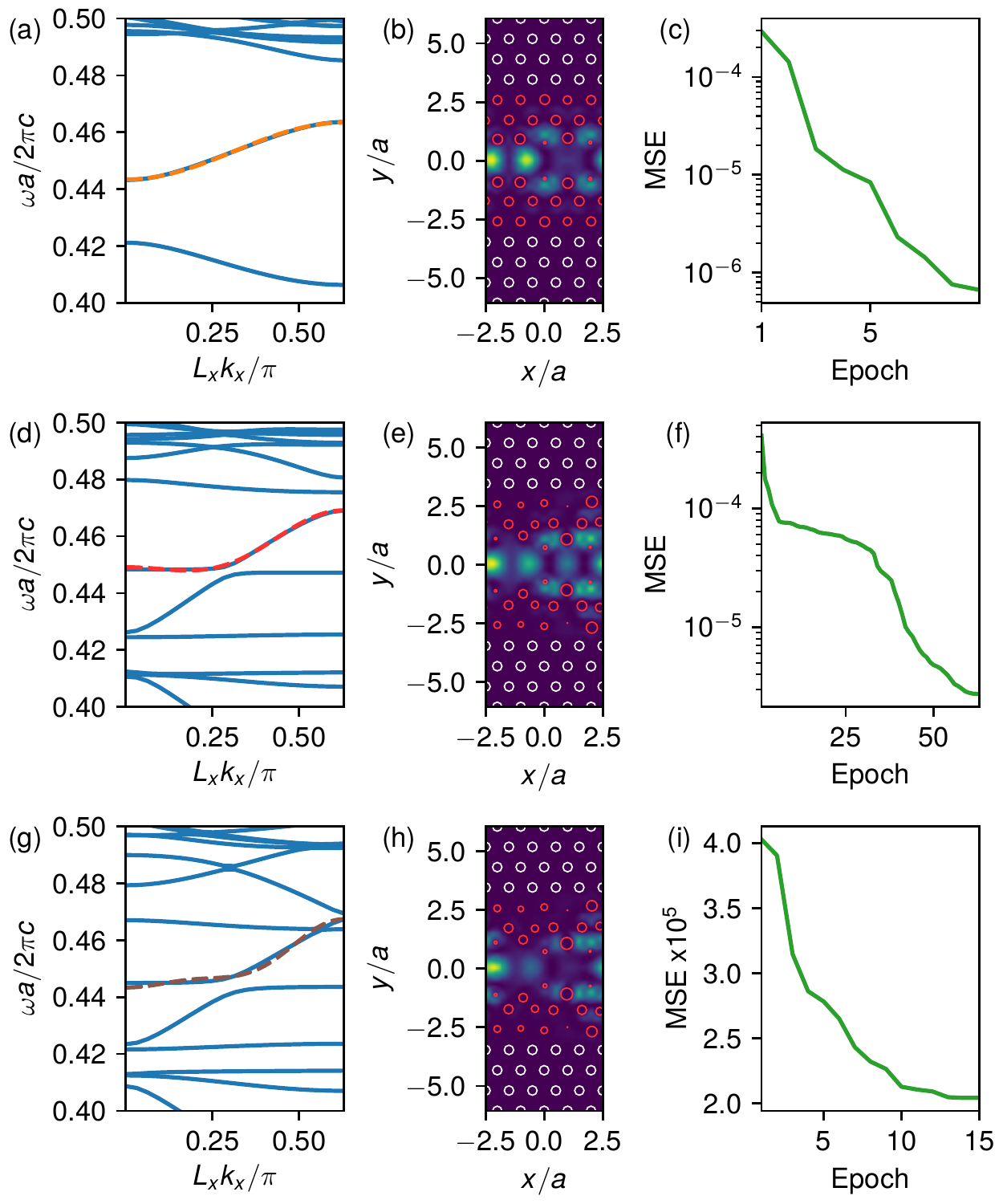}
\caption{(a): Optimized waveguide band with a target band shape given by $\bar{\omega}_1(k_x)$ in Fig. \ref{fig:wg_start}(c) (dashed line here). (b): The optimized structure and the electric field intensity of the waveguide mode at $L_x k_x/\pi = 0.5$. (c): Mean-square error vs. optimization epoch. (d)-(f): Same as (a)-(c), for the target band shape given by $\bar{\omega}_2(k_x)$ in Fig. \ref{fig:wg_start}(c). (g)-(i): Same as (a)-(c), for the target band shape given by $\bar{\omega}_3(k_x)$ in Fig. \ref{fig:wg_start}(c).}
\label{fig:wg_opt}
\end{figure}

The results of the three separate optimizations for $\bar{\omega}_i(k_x)$ are shown in Fig. \ref{fig:wg_opt}. In panel (a), we show the dispersion optimized to match $\bar{\omega}_1(k_x)$, and in panel (b) we show the corresponding final structure and the electric field intensity of one of the guided modes. We use the Limited-memory Broyden-Fletcher-Goldfarb-Shanno (LBFGS) algorithm \cite{byrd_1995,zhu_1997} to perform the optimization, and the evolution of the objective function with epochs (iterations) is shown in Fig. \ref{fig:wg_opt}(c). We note that, as discussed in Section \ref{sec:ad}, the ``backward'' propagation of the gradient with respect to \textit{all} 45 free parameters takes approximately the same time as the ``forward'' computation of the photonic bands of the structure. The optimization converges quickly to the target dispersion, with a final MSE below $10^{-6}$. 

In Fig. \ref{fig:wg_opt}(d)-(f), we show the same results, but for $\bar{\omega}_2(k_x)$. This more complicated dispersion pattern requires a greater number of optimization iterations (Fig. \ref{fig:wg_opt}(f)), but the final result also matches very well the target function, as seen in Fig. \ref{fig:wg_opt}(d). Finally, in panels (g)-(i) we show the same results for $\bar{\omega}_3(k_x)$. In the previous two optimizations, we used the un-modified structure of Fig. \ref{fig:wg_start}(d) as the initial structure. However, as the target dispersion becomes harder to achieve, this yields a poor convergence in the optimization for $\bar{\omega}_3(k_x)$. Thus, in Fig. \ref{fig:wg_opt}(g)-(i) we show the result obtained using as a starting configuration the optimal structure obtained for $\bar{\omega}_2(k_x)$, i.e. the one shown in \ref{fig:wg_opt}(e). As shown in \ref{fig:wg_opt}(g), the waveguide band does get close to the target one, but the match is not perfect. This is to be expected, as the higher-$k$ components correspond to smaller- and smaller-scale fluctuations of the real-space permittivity. This could be improved by allowing for rods of arbitrary shapes, or by increasing the supercell size $L_x$, therefore decreasing the maximum $k_x$ in units of $1/a$. 

\section{Guided-mode expansion}
\label{sec:gme}

The guided-mode expansion is a method for simulating three-dimensional (3D) layered periodic structures (PhC slabs) that, in many cases, has a computational complexity similar to the 2D plane-wave expansion, as it was formulated in Ref. \onlinecite{Andreani2006}. The method is approximate since the basis modes do not form a complete set, but it has already been shown to agree very well with first-principle methods for modes that are well-confined in the PhC region (weakly radiating in the claddings) \cite{Andreani2006, Minkov2014, Minkov2015, Minkov2018}. The key idea is to use a smart basis that captures the $z$-dependence of such modes analytically. The periodic dielectric permittivity in the $xy$-plane then enters through its Fourier components, just like in the PWE described in Section \ref{sec:pwe}. Below, we review the method in detail, and we discuss our AD implementation. As an original application, we then show an example optimization of the quality factor ($Q$) of a PhC slab cavity in a lithium niobate slab. 

\subsection{Method description}
\label{sec:gme_summary}

\begin{figure}
\centering
\includegraphics[width=0.48\textwidth]{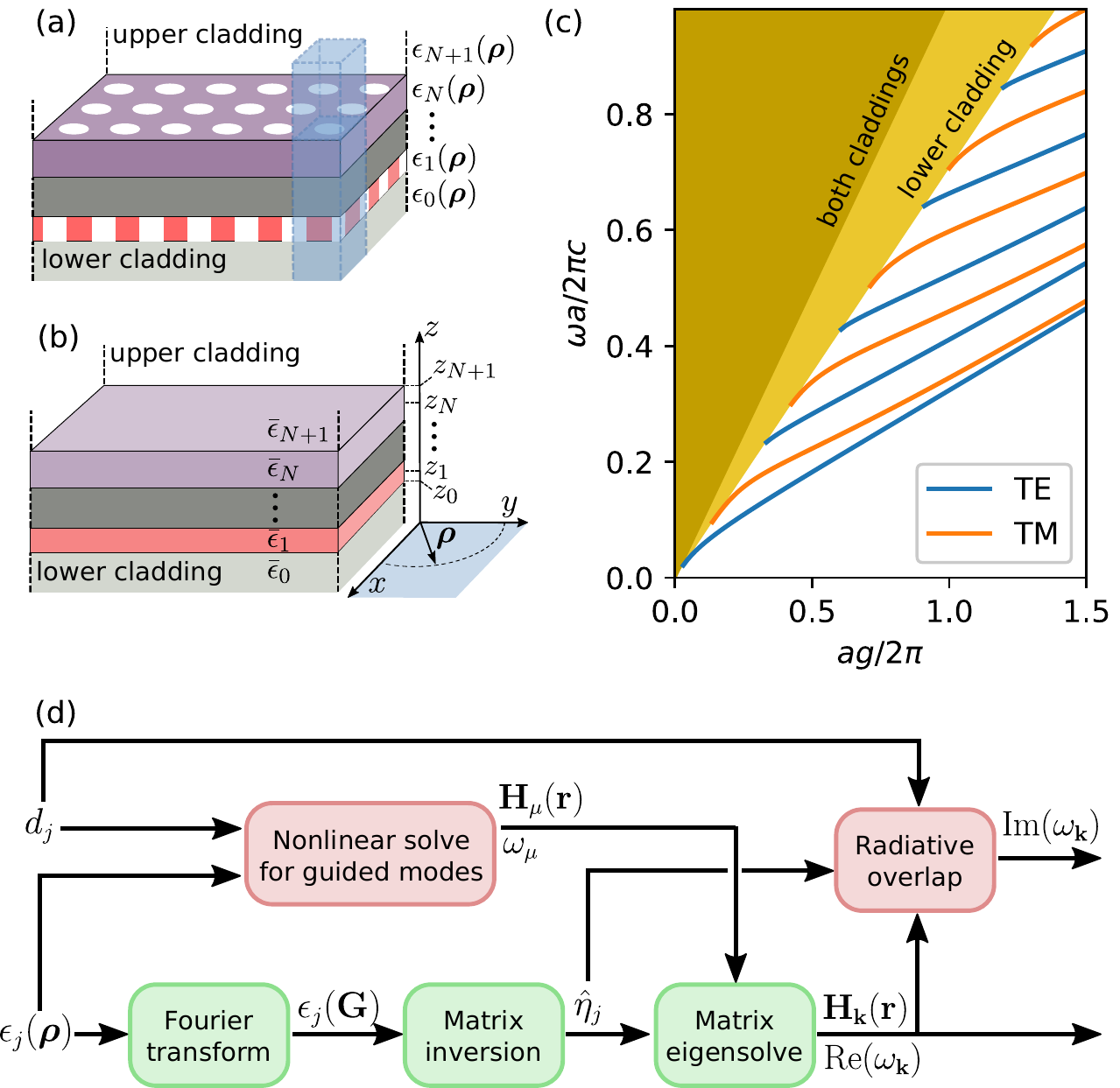}
\caption{(a): A layered photonic crystal structure consisting of lower and upper semi-infinite claddings, and $N$ layers of thickness $d_1 \dots d_N$ in between. Each layer is described by relative permittivity $\epsilon_j(\rhob)$, where $\rhob$ is the position in the $xy$-plane (cf. panel (b)). The permittivity is assumed to have the same in-plane periodicity in every layer, and the blue shaded square denotes the common elementary cell. (b): Effective homogeneous system corresponding to (a), where the permittivity $\bar{\epsilon}_j$ of each layer is the elementary-cell average of $\epsilon_j(\rhob)$. (c): Frequency $\omega$ vs. wave vector magnitude $g$ of the photonic bands of a homogeneous structure as in (b), with two layers with thickness $d_1/a = 0.3$, $d_2/a = 0.5$ ($a$ is a length unit), and permittivities $\bar{\epsilon}_0 = 2$, $\bar{\epsilon}_1 = 10$, $\bar{\epsilon}_2 = 12$, $\bar{\epsilon}_3 = 1$. Discrete guided bands can be found below the lower-cladding light line $\omega/c = g/\sqrt{2}$. Above that line, there is a continuum of modes radiating either in the lower cladding only, or in both. (d): Computational graph of the guided-mode expansion method.}
\label{fig:gme}
\end{figure}

Our implementation of the GME generalizes the seminal work of Ref. \cite{Andreani2006} to a multi-layer structure, as shown in Fig. \ref{fig:gme}(a) \cite{Zabelin2009}. Along the $z$-direction, there are two semi-infinite layers: a lower cladding extending from $z_0$ to $-\infty$, and an upper cladding extending from $z_{N+1}$ to $+\infty$. Between these two semi-infinite layers, we assume $N$ layers with thickness $d_j$. The claddings and the layers are described by in-plane permittivity distribution $\epsilon_j(\rhob)$, with $j = 0 \dots N+1$ ($j = 0$ for the lower and $j=N+1$ for the upper cladding, and vector $\rhob$ denotes the in-plane coordinate). We allow for arbitrary $\epsilon_j(\rhob)$, with the condition that the periodicity in all layers is the same, as shown by the blue shaded elementary cell in Fig. \ref{fig:gme}(a). The basis set that we use in the guided-mode expansion is then derived from a multi-layer structure with homogeneous permittivity in each layer, as shown in Fig. \ref{fig:gme}(b). Specifically, each layer has the same thickness $d_i$ as the starting structure of Fig. \ref{fig:gme}(a), and a constant relative permittivity given by
\begin{equation}
    \bar{\epsilon}_j = \frac{1}{S} \int_S \drm \rhob \epsilon_j(\rhob), \quad j = 0 \dots N+1
\end{equation}
where $S$ is the in-plane elementary cell.

As shown in Fig. \ref{fig:gme}(c), the effective homogeneous structure supports a discrete set of guided bands as well as a continuum of modes that are radiative in at least one of the claddings. This multi-layer homogeneous structure has a reflection symmetry in the $\mathbf{g}z$-plane, where $\gb$ is the in-plane wave-vector, and the modes can be separated into positive and negative eigenvalue under reflection. 
Because of the reflection symmetry in the $\mathbf{g}z$-plane, where $\gb$ is the in-plane wave-vector, the modes can again be classified as TE (electric field perpendicular to the $\mathbf{g}z$-plane, odd reflection symmetry) and TM (magnetic field perpendicular to the $\mathbf{g}z$-plane, even reflection symmetry). Notice that this classification is consistent with Section \ref{sec:pwe} and the general photonic crystal literature \cite{Joannopoulos2008}, as TE modes have $\Eb$-field in the $xy$-plane and $E_z = 0$ everywhere, while TM modes have $\Hb$-field in the $xy$-plane and $H_z = 0$ everywhere.

For any TE-polarized mode (radiative or not) at frequency $\omega$ and wave-vector $\gb$, the electric and magnetic fields in layer $j = 0 \dots N+1$ can be written most generally as
%
%\begin{widetext}
\begin{align}
    \Eb_{j} (\rhob, z) &= e^{i\gb \cdot \rhob} i \oc{\omega} \eg \left(A_{j} e^{i\chi_{j}(z-z_j)} + B_{j} e^{-i\chi_{j}(z-z_j)}\right), \label{eqn:ETE}\\
    \Hb_{j} (\rhob, z) &= e^{i\gb \cdot \rhob} i \left(A_{j}(-\chi_{j} \gh + g \zh) e^{i\chi_{j}(z-z_j)} + \right. \nonumber \\ 
    &\left. B_{j} (\chi_{j} \gh + g \zh)  e^{-i\chi_{j}(z-z_j)}\right), \label{eqn:HTE}
\end{align}
%\end{widetext}
%
where $\eg$ and $\gh$ are unit vectors pointing, respectively, in the direction orthogonal to $\gb$ and in the direction parallel to $\gb$, $g$ is the magnitude of $\gb$, and
\begin{equation}
\chi_{j} = \left(\epsb_j \oct{\omega^2} - g^2\right)^{1/2}
\label{eqn:chi}
\end{equation}
is the wave vector in the $z$-direction in layer $j$. This wave vector is either purely real (propagating fields in the layer) or purely imaginary (evanescent fields in the layer), and the square root convention is such that $\chi_{j}$ has a non-negative real or imaginary component. Furthermore, we define $z_{j}$ as the center of the $j$-th slab for $0 < j < N + 1$, and $z_0$ and $z_{N+1}$ are at the cladding interfaces (Fig. \ref{fig:gme}(b)). For TM polarization, the fields can be written as 
\begin{align}
    \Eb_{j } (\rhob, z) &= -e^{i\gb \cdot \rhob} \frac{c}{\omega} \frac{1}{\epsb_j} \left(A_{j}(-\chi_{j} \gh + g \zh) e^{i\chi_{j}(z-z_j)} + \right. \nonumber \\
    &\left. B_{j} (\chi_{j} \gh + g \zh)  e^{-i\chi_{j}(z-z_j)}\right), \label{eqn:ETM}\\
    \Hb_{j } (\rhob, z) &= e^{i\gb \cdot \rhob} \eg \left(A_{j} e^{i\chi_{j}(z-z_j)} + B_{j} e^{-i\chi_{j}(z-z_j)}\right). \label{eqn:HTM}
\end{align}
The modes also have to satisfy continuity at the layer boundaries. For TE polarization, the $H_g$ and $H_z$ components must be continuous, while for TM polarization, the continuous fields are $E_g$ and $D_z = \epsilon E_z$. 

The frequency $\omega_\mu$ and the coefficients $A_{j\mu}$ and $B_{j\mu}$ for a guided mode $\mu$ can be computed through transfer-matrix theory. The transfer matrix at an interface is determined by continuity and, for TE polarization at a frequency $\omega$, reads
\begin{equation}
    \hat{S}_j = \frac{1}{2\chi_{j+1}}
    \begin{pmatrix}
    \chi_{j+1} + \chi_{j} & \chi_{j+1} - \chi_{j} \\
    \chi_{j+1} - \chi_{j} & \chi_{j+1} + \chi_{j}
    \end{pmatrix}, 
    \label{eqn:Sj}
\end{equation}
For TM polarization, continuity imposes a similar $\hat{S}_j$ as in eq. (\ref{eqn:Sj}) above, but with the substitution $\chi_{j} \rightarrow \chi_{j}/\bar{\epsilon}_j$ everywhere. The full transfer matrix from layer $j$ to layer $j+1$ (coefficients $A_{j}$, $B_{j}$ to $A_{j+1}$, $B_{j+1}$) for both polarizations reads
\begin{align}
    \hat{T}_{j,j+1} &=
    \hat{T}_{j+1} \hat{S}_j \hat{T}_{j}, \\
    \hat{T}_{j} &= 
    \begin{pmatrix}
    e^{i\chi_{j}d_j/2} & 0\\
    0 & e^{-i\chi_{j}d_j/2} 
    \end{pmatrix},
\end{align}
where we set $d_0, d_{N+1} = 0$ for the two claddings, consistent with the definition of $z_j$ as in Fig. \ref{fig:gme}(b). The coefficients in the upper cladding are then related to the coefficients in the lower cladding by
\begin{equation}
    \begin{pmatrix}
    A_{N+1} \\
    B_{N+1}
    \end{pmatrix} =
     \hat{D}
    \begin{pmatrix}
    A_0 \\
    B_0
    \end{pmatrix} =
    \left[
    \prod_{j=0}^N \hat{T}_{j, j+1} \right]
    \begin{pmatrix}
    A_0 \\
    B_0
    \end{pmatrix}.
    \label{eqn:gme_D}
\end{equation}

A guided mode is evanescent in both claddings, which imposes $A_0 = B_{N+1} = 0$. Such guided-mode solutions can then be found numerically: we set $A_0 = 0$, $B_0 = 1$ and find $\omega_\mu$ as the solution to the equation $D_{22}(\omega) = 0$, where $D_{22}$ is the bottom right element of matrix $\hat{D}$ of eq. (\ref{eqn:gme_D}). Such modes with imaginary $\chi_0$ and $\chi_{N+1}$ (evanescent in the claddings) lie exclusively in the frequency range below the light-line of the higher-index cladding, i.e. for $\omega/c < g/\sqrt{\max(\epsb_0, \epsb_{N+1})}$, as shown in blue and orange in Fig. \ref{fig:gme}(c). Once the frequency $\omega_\mu$ of a guided mode is found, all the corresponding coefficients $\chi_{j\mu}$, $A_{j\mu}$, and $B_{j\mu}$ can also be computed. 

In the guided-mode expansion, we write an eigenmode of the PhC structure of Fig. \ref{fig:gme}(a) as
\begin{equation}
    \Hb_\kb(\rb) = \sum_\Gb \sum_{\alpha, p} c_{\Gb, \alpha, p} \Hb_{\gb, \alpha, p}^{\mathrm{guid}}(\rb),
    \label{eqn:gme_exp}
\end{equation}
where we restrict the summation only to the fully guided modes that lie below the light lines of the claddings (Fig. \ref{fig:gme}(c)). In eq. (\ref{eqn:gme_exp}), $\mathbf{g} = \Gb + \kb$, the $\Gb$-summation is over reciprocal lattice vectors just like in the plane-wave expansion of Section \ref{sec:pwe} (see Fig. \ref{fig:pwe}(c)), $\alpha$ is a band index, and $p = $ TE/TM denotes polarization. Below, we label the guided mode by a single index $\mu$ that includes $\gb, \alpha$ and $p$. Note that the two polarizations are generally mixed in the photonic crystal because of the $\rhob$-dependent permittivity, and it is not possible to classify the PhC modes strictly as TE or TM, since all of the components of both $\Eb$ and $\Hb$ can be nonzero. 

The final step in defining the expansion basis is normalizing the guided modes with respect to the scalar product for this problem, which is defined as
\begin{equation}
    (\Hb_\nu, \Hb_\mu) = \frac{1}{S}\int_S \drm \rhob \int_{-\infty}^\infty \drm z \Hb_\nu^\dagger (\rhob, z), \Hb_\mu(\rhob, z).
    \label{eqn:gme_prod}
\end{equation}
The normalization condition $(\Hb_\mu, \Hb_\mu) = 1$ for both polarizations is given in the Supplementary Information. The elements of the matrix for diagonalization $\Hcal_{\mu\nu} = (\Hb_\mu, \Th \Hb_\nu)$ can then also be computed, and are also given in the Supplementary for all the possible polarization combinations. Just like in the plane-wave expansion, the permittivity of each layer enters through the inverse matrix $\hat{\eta}_{j, \Gb, \Gb'}$ as defined in eq. (\ref{eqn:eps_ft}). The eigenvalues of $\hat{\Hcal}$ give the frequencies $\omega_\kb$ of the Bloch bands of the photonic crystal, while the eigenvectors correspond to the $c_\mu$ coefficients of eq. (\ref{eqn:gme_exp}), from which the magnetic field $\Hb_\kb$ can be computed. At this stage, the frequencies $\omega_\kb$ are purely real, because we have only included purely guided modes in the system, and $\hat{\Hcal}$ is Hermitian. The final step of the method is to compute the coupling of the photonic crystal modes to the radiative modes of the homogeneous structure that are above the light line (Fig. \ref{fig:gme}(c)).

We use perturbation theory to compute the (small) imaginary components of the frequencies of the eigenmodes of the PhC structure. These imaginary components correspond to a loss rate due to resonant coupling to radiating waves out-going in the claddings. The decay rate of an eigenvalue $\omega_\kb^2/c^2$ of matrix $\hat{\Hcal}$ can be computed in first-order time-dependent perturbation theory \cite{Ochiai2001, Andreani2006} as
\begin{equation}
    -\mathcal{I}\left(\oct{\omega_\kb^2}\right) = \pi \sum_{\Gb', p'} \sum_{o=l, u} |\mathcal{H}_{\kb r}|^2 \rho_o(\kb + \Gb', \omega_\kb),
    \label{eqn:gme_rad}
\end{equation}
where $\mathcal{H}_{\kb r}$ is a matrix element between the PhC mode and a radiative mode $\Hb_r$ given by
\begin{equation}
    \mathcal{H}_{\kb r} = \left(\Hb_k, \hat{\Theta}\Hb_r\right) = \sum_{\Gb, \alpha, p} c^*_\mu \times \left(\Hb_\mu, \hat{\Theta}\Hb_r\right).
    \label{eqn:mat_rad}
\end{equation}
For radiative modes, index $r$ includes $\Gb$, $p$, and $o$, where $o$ denotes if the mode is out-going in the lower or the upper cladding, respectively. We note again that, for guided modes, index $\mu$ includes $\Gb$, $\alpha$, and $p$. Eq. (\ref{eqn:gme_rad}) also contains the density of states of the radiative modes given by
\begin{align}
    \label{eqn:dos}
    \rho_o(\gb, \omega) &= \int_0^\infty \frac{k_z}{2\pi} \delta\left(\oct{\omega^2} - \frac{g^2 + k_z^2}{\epsb_o}\right) \\ \nonumber &= \frac{\epsb_o^{1/2}}{4\pi} \frac{\theta(\omega^2/c^2 - g^2/\epsb_o)}{(\omega^2/c^2 - g^2/\epsb_o)^{1/2}},
\end{align}
where $\theta(x)$ is the Heaviside step function. Once the imaginary part $\mathcal{I}(\omega^2/c^2)$ is computed, the imaginary part of $\omega$ can be found through $\mathcal{I}(\omega) = \mathcal{I}((\mathcal{R}(\omega^2/c^2) + i\mathcal{I}(\omega^2/c^2))^{1/2}$, and the quality factor associated to a Bloch mode $\Hb_\kb$ is $Q = \mathcal{R}(\omega_\kb)/(2\mathcal{I}(\omega_\kb))$. The normalization of the radiative modes and the matrix elements of eq. (\ref{eqn:mat_rad}) are given explicitly in the Supplementary Information.

\subsection{Automatic differentiation}
\label{sec:gme_ad}

A high-level computational graph for the GME as described in detail in Section \ref{sec:gme_summary} is presented in Fig. \ref{fig:gme}(d). In our discussions in Sections \ref{sec:ad} and \ref{sec:pwe_ad}, we have actually already covered everything that is needed for the automatic differentiation of the GME computation. The three green blocks in Fig. \ref{fig:gme}(d) are the same as those for the PWE of Fig. \ref{fig:pwe}(d). The only difference here is that the Fourier transform and matrix inversion are done for every layer $j = 0 \dots N+1$. The AD step through the nonlinear solve for the guided modes is done as defined in eq. (\ref{eqn:implicit_der}) and is the only nontrivial operation that we added to \texttt{Autograd} to enable the GME. 

As discussed in Section \ref{sec:gme_summary}, a guided-mode frequency $\omega_\mu$ is the solution to the implicit function,
\begin{equation}
    D_{22}(\omega, \epsilonb, \db) = 0,
\end{equation}
where $\epsilonb$ is a vector containing all average permittivity values $\bar{\epsilon}_j$, and $\db$ is a vector containing all layer thickness values $d_j$. This implict function is actually a simple version of eq. (\ref{eqn:implicit_der}), in that $D_{22}$ and $\omega$ are scalars and no matrix inversion is needed. Thus, once a solution $\omega_\mu$ is found, the reverse-mode AD step w.r.t. $\epsilonb$ is defined as 
\begin{equation}
    \pder{\omega_\mu}{\epsilonb} = -\left[\pder{D_{22}}{\omega_\mu}\right]^{-1} \pder{D_{22}}{\epsilonb},
    \label{eqn:D22_der}
\end{equation}
and similarly for $\db$. To implement this, we use \texttt{Autograd} to define a differentiable computation of $D_{22}$, which is computed as a product of matrices, as defined in eq. (\ref{eqn:gme_D}). We then use \texttt{Autograd} to automatically compute $\partial D_{22}/\partial\omega_\mu$, $\partial D_{22}/\partial \epsilonb$, and $\partial D_{22}/\partial \db$, using these quantities to carry out the computation defined in eq. (\ref{eqn:D22_der}) and continue the gradient accumulation through the main part of the computation.

The computation of the matrix for diagonalization from the guided modes and the Fourier components of the permittivity involves a large number of algebraic operations, which we outline in the Supplementary Information. However, these are conceptually simple to differentiate and are all supported by \texttt{Autograd}. The same is true for the radiative overlap elements, which is the final step of the forward computation shown in Fig. \ref{fig:gme}(d). This however highlights once again the importance of using automatic differentiation in handling complex computational algorithms: apart from the fundamental speed advantage offered by reverse-mode differentiation, we also profit immensely from the fact that the differentiation is split into elementary building blocks, and the extremely long and complicated derivative accumulation is done automatically. In the following section, we finally apply this to a practical problem -- optimizing a photonic crystal cavity.

\subsection{Cavity optimization}
\label{sec:gme_opt}

\begin{figure*}
\centering
\includegraphics[width=\textwidth]{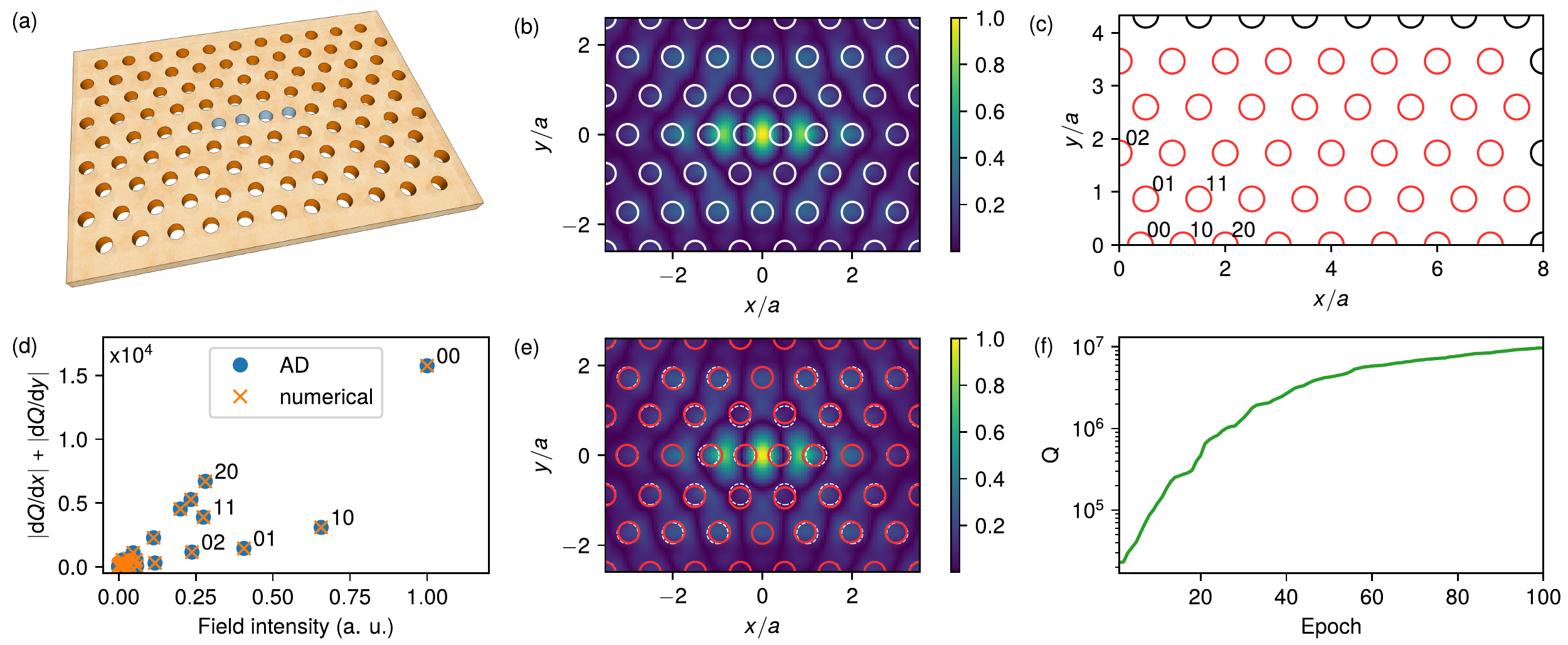}
\caption{(a): Schematic of a $L$4/3 photonic crystal cavity in a lithium niobate slab in air. The cavity is defined by four holes (marked in blue) introduced in place of the three holes of the regular lattice. (b): Electric field $|E_y|^2$ of the fundamental mode of the starting cavity. (c): One quadrant of the supercell used in the simulation. The positions of the red holes are the free parameters used in the optimization. The holes closest to the center of the cavity are labeled by their $x, y$ position index. (d): Magnitude of the gradient of the quality factor with respect to hole position vs. integrated field intensit on the surface of the corresponding hole. The labels refer to the corresponding holes in panel (c). (e): Optimized high-$Q$ cavity. The position of the holes of the starting structure are shown in white dashed lines. (f): Quality factor vs. optimization epoch.}
\label{fig:gme_opt}
\end{figure*}

Photonic crystal cavities are attractive because they can confine light to volumes close to the diffraction limit \cite{Minkov2017}. Furthermore, through small modifications of the structure, quality factors exceeding 10 million have been experimentally demonstrated in silicon PhC slabs suspended in air \cite{Asano2017}. Recently, lithium niobate (LN) has emerged as an extremely versatile material for integrated photonics applications. However, LN has a much lower refractive index than silicon ($n = 2.21$ vs. $n = 3.48$, respectively) at wavelengths around $\lambda = 1.55$\textmu m. The significantly lower index contrast with respect to the air holes and claddings makes the design of high-$Q$ PhC cavities in LN much more challenging. Recently, a design with a theoretical $Q = 1.5\times 10^6$ was proposed \cite{Li2018}, with an associated mode volume is $V = 2.43 (\lambda/n)^3$ defined as
\begin{equation}
    V = \frac{\int \drm \rb |\mathbf{E}_m(\mathbf{r})|^2 \epsilon(\mathbf{r})} {\mathrm{max}_{\mathbf{r}} \left[|\mathbf{E}_m(\mathbf{r})|^2 \epsilon(\mathbf{r})\right]}.
\end{equation}
Here, we demonstrate the optimization of a LN photonic crystal cavity with a simultaneously higher $Q$ and lower $V$ than the design in \cite{Li2018}.

The starting cavity is shown in Fig. \ref{fig:gme_opt}(a) and is based on a triangular lattice of circular holes in an LN slab. We use the $L$4/3 design \cite{Minkov2017}, in which four holes (marked in blue) are introduced in place of the three holes of the regular lattice. This type of defect results in a tightly confined cavity mode, as shown in Fig. \ref{fig:gme_opt}(b). For the underlying PhC, we use the same parameters as in Ref. \cite{Li2018}: lattice constant $a=620$nm, slab thickness $d=270$nm, and hole radius $r=145$nm. The starting cavity of Fig. \ref{fig:gme_opt}(b) has $Q = 6,100$ and a mode volume $V = 0.61 (\lambda/n)^3$. In Fig. \ref{fig:gme_opt}(c), we show one quarter of the supercell used in the simulation (the positive-$x$ and -$y$ quadrant). As optimization parameters, we use the positions of the holes marked in red, and all holes are moved symmetrically w.r.t. the $x$- and $y$-axes. Note that this also means that the holes that lie on the $x$-axis are not shifted in the $y$-direction, and vice versa for the holes on the $y$-axis. Thus, there are a total of 70 free parameters $\Delta \xb, \Delta \yb$ to optimze, and the objective function to be \textit{maximized} is simply the quality factor of the resonator, $\Lcal(\Delta \xb, \Delta \yb) = Q$.

Before optimizing, we inspect the gradient of the quality factor with respect to each hole for the starting structure. This is done both as a check that the gradients computed using automatic differentiation are correct, and to investigate the sensitivity of the $Q$ to the position of each of the holes. In Fig. \ref{fig:gme_opt}(d), we show the combined magnitude of the gradient $|\drm Q/\drm x| + |\drm Q/\drm y|$ for every hole, plotted versus the field intensity at that hole,
\begin{equation}
    I_h = \oint_h \drm \rhob |\Eb(\rhob, z_1)|^2,
\end{equation}
where the integration is along the contour of the hole, and the field $\Eb(\rhob, z_1)$ is taken at the center of the slab. As expected, there is a clear correlation between the Q-factor's sensitivity to a hole's displacement and the proximity of that hole to the center of the cavity, where the field intensity of the eigenmode is the highest. In the same panel, we also compare the AD gradient with values computed numerically using the finite-difference method, i.e. by slightly perturbing each parameter one by one. We observer that the values match perfectly. The difference is that 70 extra simulations were required to get the numerical values, while the reverse-mode AD computation after the initial $Q$ is computed took approximately the same time as the forward simulation. 

In Fig. \ref{fig:gme_opt}(e) we show the final optimized cavity, and in panel (f) we show the evolution of the quality factor with optimization steps. The LBFGS optimization is close to converged after 100 epochs, and produces a cavity mode with a GME-computed quality factor of $Q = 9.7 \times 10^6$ and a mode volume of $V = 0.49 (\lambda/n)^3$. The mode volume is smaller than that of the starting structure, because we restricted the two holes closest to the center (labeled 00 and 10 in Fig. \ref{fig:gme_opt}(c)) to only shift inwards towards the cavity center. This can be seen in panel (e), where we also show the positions of the holes for the un-optimized cavity. As is common in PhC cavity optimization and can be seen there, very small changes to the structure can result in drastic changes in the quality factor. We also computed the $Q$ of the final cavity using a first-principle finite-difference time-domain simulation in Lumerical FDTD Solutions, and found $Q = 2.4 \times 10^6$. This is slightly lower than the GME-computed result, and the discrepancy can be attributed to the fact the GME is an approximate method. In any case, this $Q$ value is sill higher than that of Ref. \cite{Li2018}, and confirms the success of the GME optimization.

\section{Discussion and conclusion}

The guided-mode expansion is an invaluable tool in the study of photonic crystal slabs. The method is approximate, but has been shown to agree well when compared to first-principle simulations in a number of different structures, while being computationally faster \cite{Andreani2006, Minkov2014, Minkov2015, Minkov2018}. Here, we used the method -- and our differentiable implementation -- to improve the quality factor of a small-volume photonic crystal cavity by more than two orders of magnitude. In the past, there have been various approaches to PhC cavity optimization. Heuristic optimizations of the quality factor have been extremely successful in silicon slabs \cite{Asano2017}, but they become challenging in lower-index materials, or when other features like the mode volume need also be taken into account. Traditional gradient-based inverse design techniques have been tried on the problem of designing high-$Q$ small-volume cavities, both using topology optimization \cite{Lu2011} and by optimizing the hole positions of a PhC resonator \cite{Wang2019}. However, in both cases the result was only moderately successful, especially with respect to the quality factor. This is fundamentally linked to the use of finite-difference frequency-domain simulations, which have significant difficulty resolving sharp resonances that shift in frequency during the course of optimization. In contrast, global optimization methods using the GME have already proven extremely useful when applied to PhC cavities \cite{Minkov2014, Minkov2017, Minkov2019}, both because of the speed of the individual computation and because high-$Q$ resonances do not pose a problem. Here, we have moved one step further, leveraging our differentiable implementation of GME to perform a gradient-based optimization. This leads to a significantly lower number of simulations needed to obtain an optimal structure. 

In periodic structures, the guided-mode expansion is particularly useful for the study of quasi-guided modes above the light line, which are hard to isolate in first-principle finite-difference methods (time- or frequency-domain). We thus anticipate that one direction in which our software package will prove invaluable is the study of bound states in the continuum \cite{Hsu2016}, which can find applications for on-chip zero-index metamaterials \cite{Minkov2018} or for enhanced nonlinear frequency conversion \cite{Minkov2019}. Another potential application is the optimization of the dispersion of a PhC slab waveguide -- similarly to what we demonstrated in Section \ref{sec:pwe_opt}, but for a fully 3D structure. This has recently been done using inverse design applied to the eigenmodes of the real-space finite-difference matrix defining Maxwell's equations \cite{Vercruysse2019}. However, this has a significant computational cost, and was furthermore only limited to modes that lie below the light line. In contrast, the GME method is extremely fast. For example, in Section \ref{sec:gme_opt}, we used a computational space of size 16x10 elementary cells, and the converged simulation takes of the order of ten minutes on a personal computer. A waveguide simulation would have a smaller supercell and thus be even faster. In addition, the optimization is not restricted to modes below the light-line, and the radiation losses can be included in the objective function, as was done for example in Ref. \cite{Minkov2015}. 

In conclusion, we have extended the paradigm of gradient-based inverse design to two mode expansion methods that are widely used in the study of photonic crystals. Namely, we have implemented the 2D plane-wave expansion and the guided-mode expansion methods in a way that allows for efficient gradient computation using reverse-mode automatic differentiation. Furthermore, this naturally brings great flexibility in defining structure parametrizations and objective functions. We have also made our software publicly available \cite{legume} for future use by the community. More broadly, our paper demonstrates the power of automatic differentiation for generic physical simulations, and highlights the fact that the sophisticated AD libraries that have been developed in the past decade \cite{autograd, paszke_pytorch, agrawal_tensorflow_2019, innes_don_2019, jax2018github} hold great promise for future large-scale inverse design of physical structures.

\section*{Acknowledgements}

This work was supported by a MURI grant from the U.S. Air Force Office of Scientific Research (FA9550-17-1-0002),  by the Department of Defense Joint Directed Energy Transition Office (DE-JTO) under Grant No. N00014-17-1-2557, and by the Gordon and Betty Moore Foundation (GBMF4744).

\bibliographystyle{achemso}
\bibliography{legume_fixed, legume_iw} 

\providecommand{\latin}[1]{#1}
\makeatletter
\providecommand{\doi}
  {\begingroup\let\do\@makeother\dospecials
  \catcode`\{=1 \catcode`\}=2 \doi@aux}
\providecommand{\doi@aux}[1]{\endgroup\texttt{#1}}
\makeatother
\providecommand*\mcitethebibliography{\thebibliography}
\csname @ifundefined\endcsname{endmcitethebibliography}
  {\let\endmcitethebibliography\endthebibliography}{}
\begin{mcitethebibliography}{63}
\providecommand*\natexlab[1]{#1}
\providecommand*\mciteSetBstSublistMode[1]{}
\providecommand*\mciteSetBstMaxWidthForm[2]{}
\providecommand*\mciteBstWouldAddEndPuncttrue
  {\def\EndOfBibitem{\unskip.}}
\providecommand*\mciteBstWouldAddEndPunctfalse
  {\let\EndOfBibitem\relax}
\providecommand*\mciteSetBstMidEndSepPunct[3]{}
\providecommand*\mciteSetBstSublistLabelBeginEnd[3]{}
\providecommand*\EndOfBibitem{}
\mciteSetBstSublistMode{f}
\mciteSetBstMaxWidthForm{subitem}{(\alph{mcitesubitemcount})}
\mciteSetBstSublistLabelBeginEnd
  {\mcitemaxwidthsubitemform\space}
  {\relax}
  {\relax}

\bibitem[Molesky \latin{et~al.}(2018)Molesky, Lin, Piggott, Jin, Vuckovic, and
  Rodriguez]{Molesky2018}
Molesky,~S.; Lin,~Z.; Piggott,~A.~Y.; Jin,~W.; Vuckovic,~J.; Rodriguez,~A.~W.
  {Outlook for inverse design in nanophotonics}. \emph{Nat. Photonics}
  \textbf{2018}, \emph{12}, 659--670\relax
\mciteBstWouldAddEndPuncttrue
\mciteSetBstMidEndSepPunct{\mcitedefaultmidpunct}
{\mcitedefaultendpunct}{\mcitedefaultseppunct}\relax
\EndOfBibitem
\bibitem[Wang \latin{et~al.}(2012)Wang, Jensen, M{\o}rk, and Sigmund]{Wang2012}
Wang,~F.; Jensen,~J.~S.; M{\o}rk,~J.; Sigmund,~O. {Systematic design of
  loss-engineered slow-light waveguides.} \emph{J. Opt. Soc. Am. A}
  \textbf{2012}, \emph{29}, 2657--66\relax
\mciteBstWouldAddEndPuncttrue
\mciteSetBstMidEndSepPunct{\mcitedefaultmidpunct}
{\mcitedefaultendpunct}{\mcitedefaultseppunct}\relax
\EndOfBibitem
\bibitem[Piggott \latin{et~al.}(2015)Piggott, Lu, Lagoudakis, Petykiewicz,
  Babinec, and Vu{\v{c}}kovi{\'{c}}]{Piggott2015}
Piggott,~A.~Y.; Lu,~J.; Lagoudakis,~K.~G.; Petykiewicz,~J.; Babinec,~T.~M.;
  Vu{\v{c}}kovi{\'{c}},~J. {Inverse design and demonstration of a compact and
  broadband on-chip wavelength demultiplexer}. \emph{Nat. Photonics}
  \textbf{2015}, \emph{9}, 374--377\relax
\mciteBstWouldAddEndPuncttrue
\mciteSetBstMidEndSepPunct{\mcitedefaultmidpunct}
{\mcitedefaultendpunct}{\mcitedefaultseppunct}\relax
\EndOfBibitem
\bibitem[Frellsen \latin{et~al.}(2016)Frellsen, Ding, Sigmund, and
  Frandsen]{Frellsen2016}
Frellsen,~L.~F.; Ding,~Y.; Sigmund,~O.; Frandsen,~L.~H. {Topology optimized
  mode multiplexing in silicon-on-insulator photonic wire waveguides}.
  \emph{Opt. Express} \textbf{2016}, \emph{24}, 16866\relax
\mciteBstWouldAddEndPuncttrue
\mciteSetBstMidEndSepPunct{\mcitedefaultmidpunct}
{\mcitedefaultendpunct}{\mcitedefaultseppunct}\relax
\EndOfBibitem
\bibitem[Lin \latin{et~al.}(2018)Lin, Christakis, Li, Mazur, Rodriguez, and
  Lon{\v{c}}ar]{Lin2018}
Lin,~Z.; Christakis,~L.; Li,~Y.; Mazur,~E.; Rodriguez,~A.~W.; Lon{\v{c}}ar,~M.
  {Topology-optimized dual-polarization Dirac cones}. \emph{Phys. Rev. B}
  \textbf{2018}, \emph{97}, 081408(R)\relax
\mciteBstWouldAddEndPuncttrue
\mciteSetBstMidEndSepPunct{\mcitedefaultmidpunct}
{\mcitedefaultendpunct}{\mcitedefaultseppunct}\relax
\EndOfBibitem
\bibitem[Lin \latin{et~al.}(2016)Lin, Pick, Lon{\v{c}}ar, and
  Rodriguez]{Lin2016}
Lin,~Z.; Pick,~A.; Lon{\v{c}}ar,~M.; Rodriguez,~A.~W. {Enhanced Spontaneous
  Emission at Third-Order Dirac Exceptional Points in Inverse-Designed Photonic
  Crystals}. \emph{Phys. Rev. Lett.} \textbf{2016}, \emph{117}, 107402\relax
\mciteBstWouldAddEndPuncttrue
\mciteSetBstMidEndSepPunct{\mcitedefaultmidpunct}
{\mcitedefaultendpunct}{\mcitedefaultseppunct}\relax
\EndOfBibitem
\bibitem[Lin \latin{et~al.}(2016)Lin, Liang, Lon{\v{c}}ar, Johnson, and
  Rodriguez]{Lin2016a}
Lin,~Z.; Liang,~X.; Lon{\v{c}}ar,~M.; Johnson,~S.~G.; Rodriguez,~A.~W.
  {Cavity-enhanced second-harmonic generation via nonlinear-overlap
  optimization}. \emph{Optica} \textbf{2016}, \emph{3}, 233\relax
\mciteBstWouldAddEndPuncttrue
\mciteSetBstMidEndSepPunct{\mcitedefaultmidpunct}
{\mcitedefaultendpunct}{\mcitedefaultseppunct}\relax
\EndOfBibitem
\bibitem[Hughes \latin{et~al.}(2018)Hughes, Minkov, Williamson, and
  Fan]{Hughes2018}
Hughes,~T.~W.; Minkov,~M.; Williamson,~I. A.~D.; Fan,~S. {Adjoint Method and
  Inverse Design for Nonlinear Nanophotonic Devices}. \emph{ACS Photonics}
  \textbf{2018}, \emph{5}, 4781--4787\relax
\mciteBstWouldAddEndPuncttrue
\mciteSetBstMidEndSepPunct{\mcitedefaultmidpunct}
{\mcitedefaultendpunct}{\mcitedefaultseppunct}\relax
\EndOfBibitem
\bibitem[Veronis \latin{et~al.}(2004)Veronis, Dutton, and Fan]{Veronis2004}
Veronis,~G.; Dutton,~R.~W.; Fan,~S. {Method for sensitivity analysis of
  photonic crystal devices}. \emph{Opt. Lett.} \textbf{2004}, \emph{29},
  2288\relax
\mciteBstWouldAddEndPuncttrue
\mciteSetBstMidEndSepPunct{\mcitedefaultmidpunct}
{\mcitedefaultendpunct}{\mcitedefaultseppunct}\relax
\EndOfBibitem
\bibitem[Jensen and Sigmund(2011)Jensen, and Sigmund]{Jensen2011}
Jensen,~J.~S.; Sigmund,~O. {Topology optimization for nano-photonics}.
  \emph{Laser Photon. Rev.} \textbf{2011}, \emph{5}, 308--321\relax
\mciteBstWouldAddEndPuncttrue
\mciteSetBstMidEndSepPunct{\mcitedefaultmidpunct}
{\mcitedefaultendpunct}{\mcitedefaultseppunct}\relax
\EndOfBibitem
\bibitem[Lalau-Keraly \latin{et~al.}(2013)Lalau-Keraly, Bhargava, Miller, and
  Yablonovitch]{Lalau-Keraly2013}
Lalau-Keraly,~C.~M.; Bhargava,~S.; Miller,~O.~D.; Yablonovitch,~E. {Adjoint
  shape optimization applied to electromagnetic design}. \emph{Opt. Express}
  \textbf{2013}, \emph{21}, 21693\relax
\mciteBstWouldAddEndPuncttrue
\mciteSetBstMidEndSepPunct{\mcitedefaultmidpunct}
{\mcitedefaultendpunct}{\mcitedefaultseppunct}\relax
\EndOfBibitem
\bibitem[Rumelhart \latin{et~al.}(1986)Rumelhart, Hinton, and
  Williams]{Rumelhart1986}
Rumelhart,~D.~E.; Hinton,~G.~E.; Williams,~R.~J. In \emph{{Parallel Distributed
  Processing}}; Rumelhart,~D.~E., McClelland,~R.~J., Eds.; MIT Press, 1986;
  Vol.~1; Chapter 8\relax
\mciteBstWouldAddEndPuncttrue
\mciteSetBstMidEndSepPunct{\mcitedefaultmidpunct}
{\mcitedefaultendpunct}{\mcitedefaultseppunct}\relax
\EndOfBibitem
\bibitem[Bischof \latin{et~al.}(1992)Bischof, Carle, Corliss, Griewank, and
  Hovland]{adifor_1992}
Bischof,~C.; Carle,~A.; Corliss,~G.; Griewank,~A.; Hovland,~P.
  ADIFOR-Generating Derivative Codes from Fortran Programs. \emph{Sci.
  Program.} \textbf{1992}, \emph{1}, 11–29\relax
\mciteBstWouldAddEndPuncttrue
\mciteSetBstMidEndSepPunct{\mcitedefaultmidpunct}
{\mcitedefaultendpunct}{\mcitedefaultseppunct}\relax
\EndOfBibitem
\bibitem[Sambridge \latin{et~al.}(2007)Sambridge, Rickwood, Rawlinson, and
  Sommacal]{sambridge_automatic_2007}
Sambridge,~M.; Rickwood,~P.; Rawlinson,~N.; Sommacal,~S. Automatic
  Differentiation in Geophysical Inverse Problems. \textbf{2007}, \emph{170},
  1--8\relax
\mciteBstWouldAddEndPuncttrue
\mciteSetBstMidEndSepPunct{\mcitedefaultmidpunct}
{\mcitedefaultendpunct}{\mcitedefaultseppunct}\relax
\EndOfBibitem
\bibitem[Enciu \latin{et~al.}(2010)Enciu, Gerbaud, and
  Wurtz]{enciu_automatic_2010}
Enciu,~P.; Gerbaud,~L.; Wurtz,~F. Automatic {{Differentiation Applied}} for
  {{Optimization}} of {{Dynamical Systems}}. \emph{IEEE Transactions on
  Magnetics} \textbf{2010}, \emph{46}, 2943--2946\relax
\mciteBstWouldAddEndPuncttrue
\mciteSetBstMidEndSepPunct{\mcitedefaultmidpunct}
{\mcitedefaultendpunct}{\mcitedefaultseppunct}\relax
\EndOfBibitem
\bibitem[Baydin \latin{et~al.}(2018)Baydin, Pearlmutter, Radul, and
  Siskind]{Baydin2018}
Baydin,~A.~G.; Pearlmutter,~B.~A.; Radul,~A.~A.; Siskind,~J.~M. {Automatic
  differentiation in machine learning: a survey}. \emph{J. Mach. Learn. Res.}
  \textbf{2018}, \emph{18}\relax
\mciteBstWouldAddEndPuncttrue
\mciteSetBstMidEndSepPunct{\mcitedefaultmidpunct}
{\mcitedefaultendpunct}{\mcitedefaultseppunct}\relax
\EndOfBibitem
\bibitem[Rackauckas \latin{et~al.}(2018)Rackauckas, Ma, Dixit, Guo, Innes,
  Revels, Nyberg, and Ivaturi]{rackauckas_comparison_2018}
Rackauckas,~C.; Ma,~Y.; Dixit,~V.; Guo,~X.; Innes,~M.; Revels,~J.; Nyberg,~J.;
  Ivaturi,~V. A {{Comparison}} of {{Automatic Differentiation}} and
  {{Continuous Sensitivity Analysis}} for {{Derivatives}} of {{Differential
  Equation Solutions}}. \emph{arXiv:1812.01892 [cs]} \textbf{2018}, \relax
\mciteBstWouldAddEndPunctfalse
\mciteSetBstMidEndSepPunct{\mcitedefaultmidpunct}
{}{\mcitedefaultseppunct}\relax
\EndOfBibitem
\bibitem[aut()]{autograd}
{Autograd: Efficiently computes derivatives of numpy code}.
  \url{https://github.com/HIPS/autograd}\relax
\mciteBstWouldAddEndPuncttrue
\mciteSetBstMidEndSepPunct{\mcitedefaultmidpunct}
{\mcitedefaultendpunct}{\mcitedefaultseppunct}\relax
\EndOfBibitem
\bibitem[Paszke \latin{et~al.}(2019)Paszke, Gross, Massa, Lerer, Bradbury,
  Chanan, Killeen, Lin, Gimelshein, Antiga, Desmaison, Kopf, Yang, DeVito,
  Raison, Tejani, Chilamkurthy, Steiner, Fang, Bai, and
  Chintala]{paszke_pytorch}
Paszke,~A. \latin{et~al.}  \emph{Advances in Neural Information Processing
  Systems 32}; 2019; pp 8024--8035\relax
\mciteBstWouldAddEndPuncttrue
\mciteSetBstMidEndSepPunct{\mcitedefaultmidpunct}
{\mcitedefaultendpunct}{\mcitedefaultseppunct}\relax
\EndOfBibitem
\bibitem[Agrawal \latin{et~al.}(2019)Agrawal, Modi, Passos, Lavoie, Agarwal,
  Shankar, Ganichev, Levenberg, Hong, Monga, and Cai]{agrawal_tensorflow_2019}
Agrawal,~A.; Modi,~A.~N.; Passos,~A.; Lavoie,~A.; Agarwal,~A.; Shankar,~A.;
  Ganichev,~I.; Levenberg,~J.; Hong,~M.; Monga,~R.; Cai,~S. {{TensorFlow
  Eager}}: {{A Multi}}-{{Stage}}, {{Python}}-{{Embedded DSL}} for {{Machine
  Learning}}. \emph{arXiv:1903.01855 [cs]} \textbf{2019}, \relax
\mciteBstWouldAddEndPunctfalse
\mciteSetBstMidEndSepPunct{\mcitedefaultmidpunct}
{}{\mcitedefaultseppunct}\relax
\EndOfBibitem
\bibitem[Innes(2019)]{innes_don_2019}
Innes,~M. Don't {{Unroll Adjoint}}: {{Differentiating SSA}}-{{Form Programs}}.
  \emph{arXiv:1810.07951 [cs]} \textbf{2019}, \relax
\mciteBstWouldAddEndPunctfalse
\mciteSetBstMidEndSepPunct{\mcitedefaultmidpunct}
{}{\mcitedefaultseppunct}\relax
\EndOfBibitem
\bibitem[Bradbury \latin{et~al.}(2018)Bradbury, Frostig, Hawkins, Johnson,
  Leary, Maclaurin, and Wanderman-Milne]{jax2018github}
Bradbury,~J.; Frostig,~R.; Hawkins,~P.; Johnson,~M.~J.; Leary,~C.;
  Maclaurin,~D.; Wanderman-Milne,~S. {JAX}: composable transformations of
  {P}ython+{N}um{P}y programs. 2018; \url{http://github.com/google/jax}\relax
\mciteBstWouldAddEndPuncttrue
\mciteSetBstMidEndSepPunct{\mcitedefaultmidpunct}
{\mcitedefaultendpunct}{\mcitedefaultseppunct}\relax
\EndOfBibitem
\bibitem[Richardson(2018)]{richardson_seismic_2018}
Richardson,~A. Seismic {{Full}}-{{Waveform Inversion Using Deep Learning
  Tools}} and {{Techniques}}. \emph{arXiv:1801.07232 [physics]} \textbf{2018},
  \relax
\mciteBstWouldAddEndPunctfalse
\mciteSetBstMidEndSepPunct{\mcitedefaultmidpunct}
{}{\mcitedefaultseppunct}\relax
\EndOfBibitem
\bibitem[Hoyer \latin{et~al.}(2019)Hoyer, {Sohl-Dickstein}, and
  Greydanus]{hoyer_neural_2019}
Hoyer,~S.; {Sohl-Dickstein},~J.; Greydanus,~S. Neural Reparameterization
  Improves Structural Optimization. \emph{arXiv:1909.04240 [cs, stat]}
  \textbf{2019}, \relax
\mciteBstWouldAddEndPunctfalse
\mciteSetBstMidEndSepPunct{\mcitedefaultmidpunct}
{}{\mcitedefaultseppunct}\relax
\EndOfBibitem
\bibitem[Rackauckas \latin{et~al.}(2020)Rackauckas, Ma, Martensen, Warner,
  Zubov, Supekar, Skinner, and Ramadhan]{rackauckas_universal_2020}
Rackauckas,~C.; Ma,~Y.; Martensen,~J.; Warner,~C.; Zubov,~K.; Supekar,~R.;
  Skinner,~D.; Ramadhan,~A. Universal {{Differential Equations}} for
  {{Scientific Machine Learning}}. \emph{arXiv:2001.04385 [cs, math, q-bio,
  stat]} \textbf{2020}, \relax
\mciteBstWouldAddEndPunctfalse
\mciteSetBstMidEndSepPunct{\mcitedefaultmidpunct}
{}{\mcitedefaultseppunct}\relax
\EndOfBibitem
\bibitem[Hughes \latin{et~al.}(2019)Hughes, Williamson, Minkov, and
  Fan]{Hughes2019}
Hughes,~T.~W.; Williamson,~I. A.~D.; Minkov,~M.; Fan,~S. {Forward-Mode
  Differentiation of Maxwell's Equations}. \emph{ACS Photonics} \textbf{2019},
  \emph{6}, 3010--3016\relax
\mciteBstWouldAddEndPuncttrue
\mciteSetBstMidEndSepPunct{\mcitedefaultmidpunct}
{\mcitedefaultendpunct}{\mcitedefaultseppunct}\relax
\EndOfBibitem
\bibitem[Hughes \latin{et~al.}(2019)Hughes, Williamson, Minkov, and
  Fan]{Hughes2019a}
Hughes,~T.~W.; Williamson,~I. A.~D.; Minkov,~M.; Fan,~S. {Wave physics as an
  analog recurrent neural network}. \emph{Sci. Adv.} \textbf{2019}, \emph{5},
  eaay6946\relax
\mciteBstWouldAddEndPuncttrue
\mciteSetBstMidEndSepPunct{\mcitedefaultmidpunct}
{\mcitedefaultendpunct}{\mcitedefaultseppunct}\relax
\EndOfBibitem
\bibitem[Su \latin{et~al.}(2019)Su, Vercruysse, Skarda, Sapra, Petykiewicz, and
  Vuckovic]{Su2019}
Su,~L.; Vercruysse,~D.; Skarda,~J.; Sapra,~N.~V.; Petykiewicz,~J.~A.;
  Vuckovic,~J. {Nanophotonic Inverse Design with SPINS: Software Architecture
  and Practical Considerations}. \emph{arXiv:1910.04829} \textbf{2019}, \relax
\mciteBstWouldAddEndPunctfalse
\mciteSetBstMidEndSepPunct{\mcitedefaultmidpunct}
{}{\mcitedefaultseppunct}\relax
\EndOfBibitem
\bibitem[Andreani and Gerace(2006)Andreani, and Gerace]{Andreani2006}
Andreani,~L.~C.; Gerace,~D. {Photonic-crystal slabs with a triangular lattice
  of triangular holes investigated using a guided-mode expansion method}.
  \emph{Phys. Rev. B} \textbf{2006}, \emph{73}, 235114\relax
\mciteBstWouldAddEndPuncttrue
\mciteSetBstMidEndSepPunct{\mcitedefaultmidpunct}
{\mcitedefaultendpunct}{\mcitedefaultseppunct}\relax
\EndOfBibitem
\bibitem[leg()]{legume}
{Legume: Guided Mode Expansion supporting automatic differentiation with
  autograd}. \url{https://github.com/fancompute/legume}\relax
\mciteBstWouldAddEndPuncttrue
\mciteSetBstMidEndSepPunct{\mcitedefaultmidpunct}
{\mcitedefaultendpunct}{\mcitedefaultseppunct}\relax
\EndOfBibitem
\bibitem[Boyd(2003)]{Boyd2003}
Boyd,~R.~W. \emph{{Nonlinear optics}}; Elsevier, 2003\relax
\mciteBstWouldAddEndPuncttrue
\mciteSetBstMidEndSepPunct{\mcitedefaultmidpunct}
{\mcitedefaultendpunct}{\mcitedefaultseppunct}\relax
\EndOfBibitem
\bibitem[Baba(2008)]{Baba2008}
Baba,~T. {Slow light in photonic crystals}. \emph{Nat. Photonics}
  \textbf{2008}, \emph{2}, 465--473\relax
\mciteBstWouldAddEndPuncttrue
\mciteSetBstMidEndSepPunct{\mcitedefaultmidpunct}
{\mcitedefaultendpunct}{\mcitedefaultseppunct}\relax
\EndOfBibitem
\bibitem[Arizmendi(2004)]{Arizmendi2004}
Arizmendi,~L. {Photonic applications of lithium niobate crystals}. \emph{Phys.
  status solidi} \textbf{2004}, \emph{201}, 253--283\relax
\mciteBstWouldAddEndPuncttrue
\mciteSetBstMidEndSepPunct{\mcitedefaultmidpunct}
{\mcitedefaultendpunct}{\mcitedefaultseppunct}\relax
\EndOfBibitem
\bibitem[Wengert(1964)]{Wengert1964}
Wengert,~R.~E. {A Simple Automatic Derivative Evaluation Program}.
  \emph{Commun. ACM} \textbf{1964}, \emph{7}, 463--464\relax
\mciteBstWouldAddEndPuncttrue
\mciteSetBstMidEndSepPunct{\mcitedefaultmidpunct}
{\mcitedefaultendpunct}{\mcitedefaultseppunct}\relax
\EndOfBibitem
\bibitem[Griewank and Walther(2008)Griewank, and Walther]{Griewank2008}
Griewank,~A.; Walther,~A. \emph{{Evaluating derivatives: principles and
  techniques of algorithmic differentiation}}; Siam, 2008; Vol. 105\relax
\mciteBstWouldAddEndPuncttrue
\mciteSetBstMidEndSepPunct{\mcitedefaultmidpunct}
{\mcitedefaultendpunct}{\mcitedefaultseppunct}\relax
\EndOfBibitem
\bibitem[Dauvergne and Hasco{\"{e}}t(2006)Dauvergne, and
  Hasco{\"{e}}t]{Dauvergne2006}
Dauvergne,~B.; Hasco{\"{e}}t,~L. {The data-flow equations of checkpointing in
  reverse automatic differentiation}. Int. Conf. Comput. Sci. 2006; pp
  566--573\relax
\mciteBstWouldAddEndPuncttrue
\mciteSetBstMidEndSepPunct{\mcitedefaultmidpunct}
{\mcitedefaultendpunct}{\mcitedefaultseppunct}\relax
\EndOfBibitem
\bibitem[{van der Walt} \latin{et~al.}(2011){van der Walt}, {Colbert}, and
  {Varoquaux}]{numpy_2011}
{van der Walt},~S.; {Colbert},~S.~C.; {Varoquaux},~G. The NumPy Array: A
  Structure for Efficient Numerical Computation. \emph{Computing in Science
  Engineering} \textbf{2011}, \emph{13}, 22--30\relax
\mciteBstWouldAddEndPuncttrue
\mciteSetBstMidEndSepPunct{\mcitedefaultmidpunct}
{\mcitedefaultendpunct}{\mcitedefaultseppunct}\relax
\EndOfBibitem
\bibitem[Virtanen \latin{et~al.}(2020)Virtanen, Gommers, Oliphant, Haberland,
  Reddy, Cournapeau, Burovski, Peterson, Weckesser, Bright, van~der Walt,
  Brett, Wilson, Millman, Mayorov, Nelson, Jones, Kern, Larson, Carey, Polat,
  Feng, Moore, VanderPlas, Laxalde, Perktold, Cimrman, Henriksen, Quintero,
  Harris, Archibald, Ribeiro, Pedregosa, and van Mulbregt]{virtanen_scipy_2020}
Virtanen,~P. \latin{et~al.}  {{SciPy}} 1.0: Fundamental Algorithms for
  Scientific Computing in {{Python}}. \emph{Nature Methods} \textbf{2020},
  1--12\relax
\mciteBstWouldAddEndPuncttrue
\mciteSetBstMidEndSepPunct{\mcitedefaultmidpunct}
{\mcitedefaultendpunct}{\mcitedefaultseppunct}\relax
\EndOfBibitem
\bibitem[Joannopoulos \latin{et~al.}(2008)Joannopoulos, Johnson, Winn, and
  Others]{Joannopoulos2008}
Joannopoulos,~J.~D.; Johnson,~S.~G.; Winn,~J.~N.; Others, \emph{{Photonic
  Crystals: Molding the Flow of Light}}; Princeton University Press, 2008\relax
\mciteBstWouldAddEndPuncttrue
\mciteSetBstMidEndSepPunct{\mcitedefaultmidpunct}
{\mcitedefaultendpunct}{\mcitedefaultseppunct}\relax
\EndOfBibitem
\bibitem[Lalanne \latin{et~al.}(2018)Lalanne, Yan, Vynck, Sauvan, and
  Hugonin]{Lalanne2018}
Lalanne,~P.; Yan,~W.; Vynck,~K.; Sauvan,~C.; Hugonin,~J.~P. {Light Interaction
  with Photonic and Plasmonic Resonances}. \emph{Laser Photonics Rev.}
  \textbf{2018}, \emph{12}, 1--38\relax
\mciteBstWouldAddEndPuncttrue
\mciteSetBstMidEndSepPunct{\mcitedefaultmidpunct}
{\mcitedefaultendpunct}{\mcitedefaultseppunct}\relax
\EndOfBibitem
\bibitem[Minkov and Savona(2014)Minkov, and Savona]{Minkov2014}
Minkov,~M.; Savona,~V. {Automated optimization of photonic crystal slab
  cavities}. \emph{Sci. Rep.} \textbf{2014}, \emph{4}, 5124\relax
\mciteBstWouldAddEndPuncttrue
\mciteSetBstMidEndSepPunct{\mcitedefaultmidpunct}
{\mcitedefaultendpunct}{\mcitedefaultseppunct}\relax
\EndOfBibitem
\bibitem[Minkov and Savona(2015)Minkov, and Savona]{Minkov2015}
Minkov,~M.; Savona,~V. {Wide-band slow light in compact photonic crystal
  coupled-cavity waveguides}. \emph{Optica} \textbf{2015}, \emph{2},
  631--634\relax
\mciteBstWouldAddEndPuncttrue
\mciteSetBstMidEndSepPunct{\mcitedefaultmidpunct}
{\mcitedefaultendpunct}{\mcitedefaultseppunct}\relax
\EndOfBibitem
\bibitem[Minkov \latin{et~al.}(2018)Minkov, Williamson, Xiao, and
  Fan]{Minkov2018}
Minkov,~M.; Williamson,~I. A.~D.; Xiao,~M.; Fan,~S. {Zero-index bound states in
  the continuum}. \emph{Phys. Rev. Lett.} \textbf{2018}, \emph{121},
  263901\relax
\mciteBstWouldAddEndPuncttrue
\mciteSetBstMidEndSepPunct{\mcitedefaultmidpunct}
{\mcitedefaultendpunct}{\mcitedefaultseppunct}\relax
\EndOfBibitem
\bibitem[Ho \latin{et~al.}(1990)Ho, Chan, and Soukoulis]{Ho1990}
Ho,~K.~M.; Chan,~C.~T.; Soukoulis,~C.~M. {Existence of a photonic gap in
  periodic dielectric structures}. \emph{Phys. Rev. Lett.} \textbf{1990},
  \emph{65}, 3152\relax
\mciteBstWouldAddEndPuncttrue
\mciteSetBstMidEndSepPunct{\mcitedefaultmidpunct}
{\mcitedefaultendpunct}{\mcitedefaultseppunct}\relax
\EndOfBibitem
\bibitem[Li(1996)]{Li1996}
Li,~L. {Use of Fourier series in the analysis of discontinuous periodic
  structures}. \emph{J. Opt. Soc. Am. A} \textbf{1996}, \emph{13},
  1870--1876\relax
\mciteBstWouldAddEndPuncttrue
\mciteSetBstMidEndSepPunct{\mcitedefaultmidpunct}
{\mcitedefaultendpunct}{\mcitedefaultseppunct}\relax
\EndOfBibitem
\bibitem[Giles(2008)]{Giles2008}
Giles,~M.~B. {Collected matrix derivative results for forward and reverse mode
  algorithmic differentiation}. \emph{Lect. Notes Comput. Sci. Eng.}
  \textbf{2008}, \emph{64 LNCSE}, 35--44\relax
\mciteBstWouldAddEndPuncttrue
\mciteSetBstMidEndSepPunct{\mcitedefaultmidpunct}
{\mcitedefaultendpunct}{\mcitedefaultseppunct}\relax
\EndOfBibitem
\bibitem[Lee(2007)]{Lee2007}
Lee,~T.~H. {Adjoint method for design sensitivity analysis of multiple
  eigenvalues and associated eigenvectors}. \emph{AIAA J.} \textbf{2007},
  \emph{45}, 1998--2004\relax
\mciteBstWouldAddEndPuncttrue
\mciteSetBstMidEndSepPunct{\mcitedefaultmidpunct}
{\mcitedefaultendpunct}{\mcitedefaultseppunct}\relax
\EndOfBibitem
\bibitem[Johnson and Joannopoulos(2001)Johnson, and Joannopoulos]{Johnson2001}
Johnson,~S.; Joannopoulos,~J. {Block-iterative frequency-domain methods for
  Maxwell's equations in a planewave basis}. \emph{Opt. Express} \textbf{2001},
  \emph{8}, 173\relax
\mciteBstWouldAddEndPuncttrue
\mciteSetBstMidEndSepPunct{\mcitedefaultmidpunct}
{\mcitedefaultendpunct}{\mcitedefaultseppunct}\relax
\EndOfBibitem
\bibitem[Lee(1983)]{Lee1983}
Lee,~S.-w. {Fourier Transform of a Polygonal Shape Function and Its Application
  in Electromagnetics}. \emph{IEEE Trans. Antennas Propag.} \textbf{1983},
  \emph{AP-31}, 99--103\relax
\mciteBstWouldAddEndPuncttrue
\mciteSetBstMidEndSepPunct{\mcitedefaultmidpunct}
{\mcitedefaultendpunct}{\mcitedefaultseppunct}\relax
\EndOfBibitem
\bibitem[Moss \latin{et~al.}(2013)Moss, Morandotti, Gaeta, and
  Lipson]{Moss2013}
Moss,~D.~J.; Morandotti,~R.; Gaeta,~A.~L.; Lipson,~M. {New CMOS-compatible
  platforms based on silicon nitride and Hydex for nonlinear optics}.
  \emph{Nat. Photonics} \textbf{2013}, \emph{7}, 597\relax
\mciteBstWouldAddEndPuncttrue
\mciteSetBstMidEndSepPunct{\mcitedefaultmidpunct}
{\mcitedefaultendpunct}{\mcitedefaultseppunct}\relax
\EndOfBibitem
\bibitem[Byrd \latin{et~al.}(1995)Byrd, Lu, Nocedal, and Zhu]{byrd_1995}
Byrd,~R.~H.; Lu,~P.; Nocedal,~J.; Zhu,~C. A Limited Memory Algorithm for Bound
  Constrained Optimization. \emph{SIAM Journal on Scientific Computing}
  \textbf{1995}, \emph{16}, 1190--1208\relax
\mciteBstWouldAddEndPuncttrue
\mciteSetBstMidEndSepPunct{\mcitedefaultmidpunct}
{\mcitedefaultendpunct}{\mcitedefaultseppunct}\relax
\EndOfBibitem
\bibitem[Zhu \latin{et~al.}(1997)Zhu, Byrd, Lu, and Nocedal]{zhu_1997}
Zhu,~C.; Byrd,~R.~H.; Lu,~P.; Nocedal,~J. Algorithm 778: L-BFGS-B: Fortran
  Subroutines for Large-Scale Bound-Constrained Optimization. \emph{ACM Trans.
  Math. Softw.} \textbf{1997}, \emph{23}, 550–560\relax
\mciteBstWouldAddEndPuncttrue
\mciteSetBstMidEndSepPunct{\mcitedefaultmidpunct}
{\mcitedefaultendpunct}{\mcitedefaultseppunct}\relax
\EndOfBibitem
\bibitem[Zabelin(2009)]{Zabelin2009}
Zabelin,~V. {Numerical Investigations of Two-Dimensional Photonic Crystal
  Optical Properties, Design and Analysis of Photonic Crystal Based
  Structures}. Ph.D.\ thesis, 2009\relax
\mciteBstWouldAddEndPuncttrue
\mciteSetBstMidEndSepPunct{\mcitedefaultmidpunct}
{\mcitedefaultendpunct}{\mcitedefaultseppunct}\relax
\EndOfBibitem
\bibitem[Minkov \latin{et~al.}(2017)Minkov, Savona, and Gerace]{Minkov2017}
Minkov,~M.; Savona,~V.; Gerace,~D. {Photonic crystal slab cavity simultaneously
  optimized for ultra-high Q / v and vertical radiation coupling}. \emph{Appl.
  Phys. Lett.} \textbf{2017}, \emph{111}, 131104\relax
\mciteBstWouldAddEndPuncttrue
\mciteSetBstMidEndSepPunct{\mcitedefaultmidpunct}
{\mcitedefaultendpunct}{\mcitedefaultseppunct}\relax
\EndOfBibitem
\bibitem[Asano \latin{et~al.}(2017)Asano, Ochi, Takahashi, Kishimoto, and
  Noda]{Asano2017}
Asano,~T.; Ochi,~Y.; Takahashi,~Y.; Kishimoto,~K.; Noda,~S. {Photonic crystal
  nanocavity with a Q factor exceeding eleven million}. \emph{Opt. Express}
  \textbf{2017}, \emph{25}, 1769--1777\relax
\mciteBstWouldAddEndPuncttrue
\mciteSetBstMidEndSepPunct{\mcitedefaultmidpunct}
{\mcitedefaultendpunct}{\mcitedefaultseppunct}\relax
\EndOfBibitem
\bibitem[Li \latin{et~al.}(2018)Li, Liang, Luo, He, and Lin]{Li2018}
Li,~M.; Liang,~H.; Luo,~R.; He,~Y.; Lin,~Q. {High-Q two-dimensional lithium
  niobate photonic crystal slab nanoresonators}. \emph{arXiv:1806.04755}
  \textbf{2018}, \relax
\mciteBstWouldAddEndPunctfalse
\mciteSetBstMidEndSepPunct{\mcitedefaultmidpunct}
{}{\mcitedefaultseppunct}\relax
\EndOfBibitem
\bibitem[Lu \latin{et~al.}(2011)Lu, Boyd, and Vu{\v{c}}kovi{\'{c}}]{Lu2011}
Lu,~J.; Boyd,~S.; Vu{\v{c}}kovi{\'{c}},~J. {Inverse design of a
  three-dimensional nanophotonic resonator}. \emph{Opt. Express} \textbf{2011},
  \emph{19}, 10563--10570\relax
\mciteBstWouldAddEndPuncttrue
\mciteSetBstMidEndSepPunct{\mcitedefaultmidpunct}
{\mcitedefaultendpunct}{\mcitedefaultseppunct}\relax
\EndOfBibitem
\bibitem[Wang \latin{et~al.}(2019)Wang, Zheng, Decker, Wu, Essertel, and
  Rompf]{Wang2019}
Wang,~F.; Zheng,~D.; Decker,~J.; Wu,~X.; Essertel,~G.~M.; Rompf,~T.
  {Demystifying differentiable programming: shift/reset the penultimate
  backpropagator}. \emph{Proc. ACM Program. Lang.} \textbf{2019}, \emph{3},
  1--31\relax
\mciteBstWouldAddEndPuncttrue
\mciteSetBstMidEndSepPunct{\mcitedefaultmidpunct}
{\mcitedefaultendpunct}{\mcitedefaultseppunct}\relax
\EndOfBibitem
\bibitem[Minkov \latin{et~al.}(2019)Minkov, Gerace, and Fan]{Minkov2019}
Minkov,~M.; Gerace,~D.; Fan,~S. {Doubly resonant $\chi$ (2) nonlinear photonic
  crystal cavity based on a bound state in the continuum}. \emph{Optica}
  \textbf{2019}, \emph{6}, 1039--1045\relax
\mciteBstWouldAddEndPuncttrue
\mciteSetBstMidEndSepPunct{\mcitedefaultmidpunct}
{\mcitedefaultendpunct}{\mcitedefaultseppunct}\relax
\EndOfBibitem
\bibitem[Hsu \latin{et~al.}(2016)Hsu, Zhen, Stone, Joannopoulos, and
  Soljacic]{Hsu2016}
Hsu,~C.~W.; Zhen,~B.; Stone,~A.~D.; Joannopoulos,~J.~D.; Soljacic,~M. {Bound
  states in the continuum}. \emph{Nat. Rev. Mater.} \textbf{2016}, \emph{1},
  16048\relax
\mciteBstWouldAddEndPuncttrue
\mciteSetBstMidEndSepPunct{\mcitedefaultmidpunct}
{\mcitedefaultendpunct}{\mcitedefaultseppunct}\relax
\EndOfBibitem
\bibitem[Vercruysse \latin{et~al.}(2019)Vercruysse, Sapra, Su, Trivedi, and
  Vu{\v{c}}kovi{\'{c}}]{Vercruysse2019}
Vercruysse,~D.; Sapra,~N.~V.; Su,~L.; Trivedi,~R.; Vu{\v{c}}kovi{\'{c}},~J.
  {Analytical level set fabrication constraints for inverse design}. \emph{Sci.
  Rep.} \textbf{2019}, \emph{9}, 8999\relax
\mciteBstWouldAddEndPuncttrue
\mciteSetBstMidEndSepPunct{\mcitedefaultmidpunct}
{\mcitedefaultendpunct}{\mcitedefaultseppunct}\relax
\EndOfBibitem
\bibitem[Ochiai and Sakoda(2001)Ochiai, and Sakoda]{Ochiai01}
Ochiai,~T.; Sakoda,~K. {Nearly free-photon approximation for two-dimensional
  photonic crystal slabs}. \emph{Phys. Rev. B} \textbf{2001}, \emph{64},
  45108\relax
\mciteBstWouldAddEndPuncttrue
\mciteSetBstMidEndSepPunct{\mcitedefaultmidpunct}
{\mcitedefaultendpunct}{\mcitedefaultseppunct}\relax
\EndOfBibitem
\bibitem[Carniglia and Mandel(1971)Carniglia, and Mandel]{Carniglia1971}
Carniglia,~C.~K.; Mandel,~L. {Quantization of Evanescent Electromagnetic
  Waves}. \emph{Phys. Rev. D} \textbf{1971}, \emph{3}, 280--296\relax
\mciteBstWouldAddEndPuncttrue
\mciteSetBstMidEndSepPunct{\mcitedefaultmidpunct}
{\mcitedefaultendpunct}{\mcitedefaultseppunct}\relax
\EndOfBibitem
\end{mcitethebibliography}

\onecolumngrid
\clearpage

\renewcommand{\thefigure}{S\arabic{figure}}
\setcounter{figure}{0}
\renewcommand{\thetable}{S\arabic{table}} 
\setcounter{table}{0}
\renewcommand{\theequation}{S\arabic{equation}} 
\setcounter{equation}{0}
\renewcommand{\thesection}{S\arabic{section}} 
\setcounter{section}{0}

\section*{Inverse design of photonic crystals through automatic differentiation: supplementary information}

Here, we provide some mathematical details related to the guided-mode expansion. Equation and citation references are with respect to the main text. 

\section{Guided-mode normalization}

The integration along the $z$-direction in eq. (\ref{eqn:gme_prod}) for fields of the form of (\ref{eqn:HTE}) and (\ref{eqn:HTM}) can be handled analytically. For this purpose, we define 
\begin{equation}
    I_j(a) = \begin{cases}
    \int_{-\infty}^0 \drm z e^{iaz} = -\frac{i}{a} & $j = 0$,\;\mathcal{I}(a) \le 0 \\[5pt]
      \int_0^\infty \drm z e^{iaz} = \frac{i}{a} & $j = N+1$,\;\mathcal{I}(a) \ge 0 \\[5pt]
    \int_{-d_j/2}^{d_j/2} \drm z e^{ia z} = \frac{2}{a} \sin\left(\frac{a d_j}{2}\right) & 1 \le j \le N
    \end{cases}
\end{equation}

For the normalization $(\Hb_\mu, \Hb_\mu) = 1$ of the TE guided modes, using eq. (\ref{eqn:HTE}) and omitting index $\mu$ for brevity, we then find:
\begin{align}
1 = \sum_{j=0}^{N+1}&\left[ (|\chi_j|^2 + g^2) (|A_j|^2 I_j(-\chi_j^* + \chi_j) + |B_j|^2 I_j(\chi_j^* - \chi_j)) + \right. \label{eqn:TEnorm} \\ \nonumber &\left. (g^2 - |\chi_j|^2)(A_j^* B_j I_j(-\chi_j^* - \chi_j) + B_j^* A_j I_j(\chi_j^* + \chi_j)) \right],
\end{align}

For TM polarization, using eq. (\ref{eqn:HTM}), we find
\begin{align}
 \label{eqn:TMnorm} 
1 = \sum_{j=0}^{N+1}\left[|A_j|^2 I_j(-\chi_j^* + \chi_j) + |B_j|^2 I_j(\chi_j^* - \chi_j) + A_j^* B_j I_j(-\chi_j^* - \chi_j) + B_j^* A_j I_j(\chi_j^* + \chi_j) \right],
\end{align}
Thus, after computing the coefficients $A_{j\mu}, B_{j\mu}$ for a guided mode, we rescale them by the square root of the terms on the righ-hand side in eqs. (\ref{eqn:TEnorm}) and (\ref{eqn:TMnorm}) to impose the normalization.

\section{Radiative mode normalization}

The radiative modes that have a propagating component in at least one of the claddings must be treated with special care. First, we introduce a normalization box extending from $-L$ to $L$ in the $z$-direction and modify Eq.~(\ref{eqn:gme_prod}) into:
\begin{equation}
    (\Hb_r, \Hb_r) = \frac{1}{S} \int_S \drm \rhob
    \lim_{L\rightarrow\infty}\frac{1}{L}\int_{-L}^{L} \drm z \Hb_r^\dagger \Hb_r.
\label{eqn:norm_rad}
\end{equation}
As $L$ increases, the contribution to the integral from the inner layers becomes negligible and only the cladding layers contribute. 
Second, we notice that radiative modes used to calculate diffraction losses should satisfy the outgoing wave boundary condition, as the outgoing components are the ones that determine the energy flux far from the waveguide. 
For this reason, only the outgoing components contribute to the normalization of the radiative modes.
As discussed in Ref.~\cite{Ochiai01}, this is formally obtained by adding a small negative imaginary part to the squared frequency, which is set to zero at the end of the calculation. Notice that this imaginary part of the squared frequency is consistent with the theory of Quasi Normal Modes, which also satisfy outgoing wave boundary conditions \cite{Lalanne2018}.
With these prescriptions, we get 
\begin{equation}
    (\Hb_r, \Hb_r) = \epsb_0 \oct{\omega^2} |B_0|^2 + \epsb_{N+1} \oct{\omega^2} |A_{N+1}|^2
    \label{eqn:te_rad_norm}
\end{equation}
for TE polarization, and
\begin{equation}
    (\Hb_r, \Hb_r) =  |B_0|^2 + |A_{N+1}|^2
   \label{eqn:tm_rad_norm}
\end{equation}
for TM polarization. If instead the normalization is kept as in Eq.~(\ref{eqn:gme_prod})
(i.e., the factor $1/L$ is dropped in Eq.~(\ref{eqn:norm_rad})), expressions 
(\ref{eqn:te_rad_norm})-(\ref{eqn:tm_rad_norm}) would be $\propto L$, the matrix elements between a guided and a radiative mode would be $\propto 1/\sqrt{L}$, the density of states in Eq.~(\ref{eqn:dos}) would have an additional factor of $L$, and the final results for the intrinsic losses would be the same.

\begin{figure*}[h]
\centering
\vspace*{-0cm}
\hspace*{1cm}
\includegraphics[width=\textwidth]{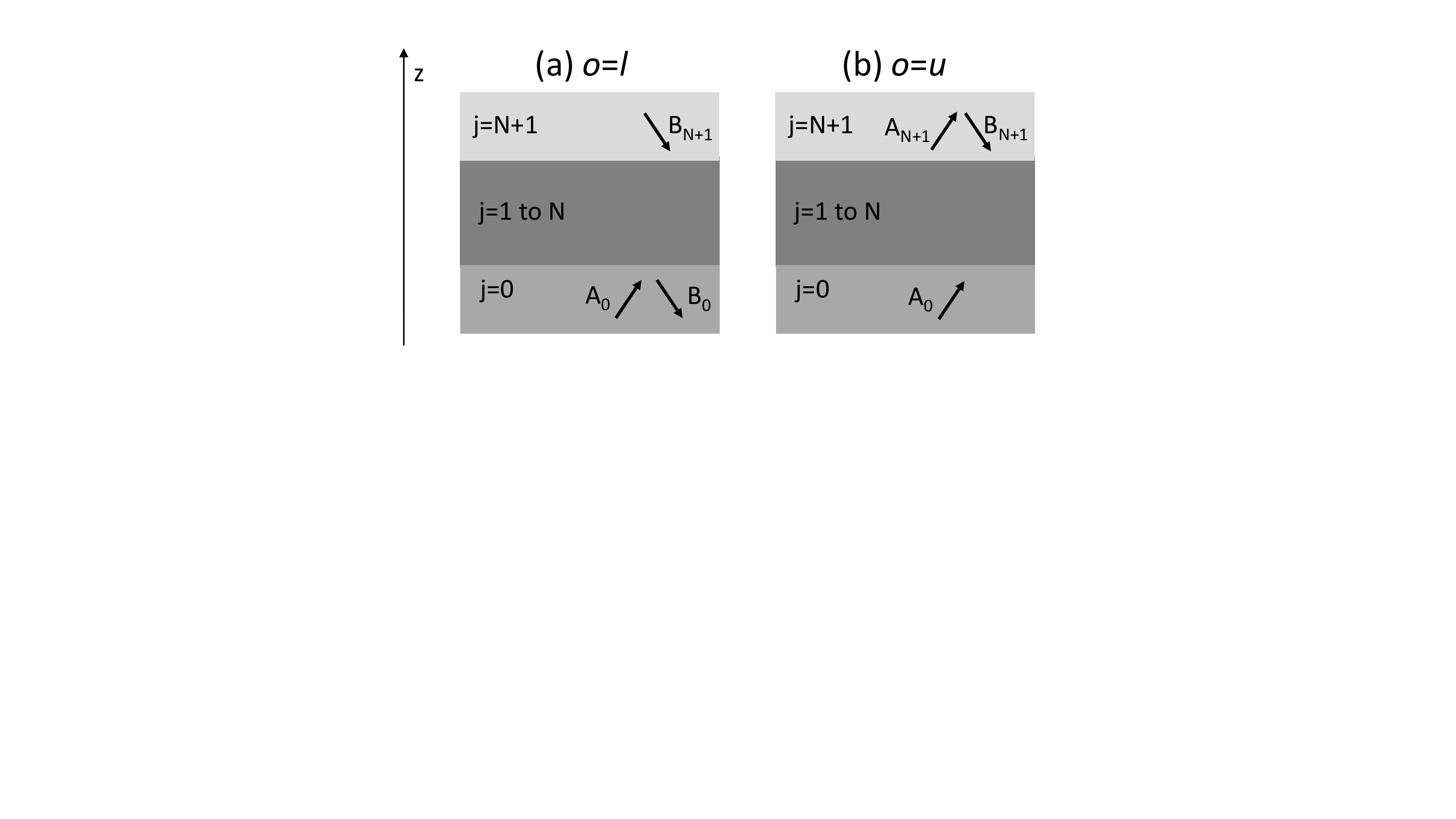}
\vspace*{-6cm}
\caption{Schematic representation of radiative states that are outgoing (a) in the lower cladding, (b) in the upper cladding. Only the outgoing component ($B_0$ in (a) and $A_{N+1}$ in (b) contributes to the normalization.}
\label{fig:outgoing}
\end{figure*}

In order to specify the states to be used in the application of first-order perturbation theory, we recall that the radiative modes of the effective waveguide can be expressed in terms of basis states with an outgoing component in either the lower or the upper cladding \cite{Carniglia1971,Andreani2006}, see schematic representation in Fig.~\ref{fig:outgoing}.
The state that is outgoing in the lower cladding ($o=l$ in Eq.~(\ref{eqn:dos})) has  $A_{N+1}=0$ and the condition of unit normalization imposes $B_0=\frac{c}{\omega\sqrt{\epsb_0}}$ for TE modes, $B_0=1$ for TM modes.
The state that is outgoing in the upper cladding ($o=u$ in Eq.~(\ref{eqn:dos})) has  $B_{0}=0$ and the condition of unit normalization imposes $A_{N+1}=\frac{c}{\omega\sqrt{\epsb_{N+1}}}$ for TE modes, $A_{N+1}=1$ for TM modes.

A special situation occurs when the lower and upper claddings have different dielectric constants, $\epsb_{0}\ne\epsb_{N+1}$,
and the frequency of a mode lies between the light lines of the lower and upper claddings. In this case, one of the two wavevectors $\chi_0$, $\chi_{N+1}$ of Eq.~(\ref{eqn:chi}) is real while the other one is imaginary, and the mode is semi-radiative, as it radiates energy in one cladding only. In this situation, only the coefficient belonging to the radiative component appears in the normalizations (\ref{eqn:te_rad_norm})-(\ref{eqn:tm_rad_norm}). Moreover, when applying perturbation theory in Eq.~(\ref{eqn:gme_rad}), only the index $o$ corresponding to the radiative component has to be kept. Notice that the density of states in Eq.~(\ref{eqn:dos}) is automatically zero for the evanescent component, thanks to the Heaviside function.

\section{Matrix elements}

The scalar product for the matrix elements can be transformed to \cite{Andreani2006}
\begin{equation}
    (\Hb_\mu, \Th \Hb_\nu) = \frac{1}{S} \int_S \drm \rhob \int_{-\infty}^\infty \drm z \frac{1}{\epsilon(\rb)} (\nabla \times \Hb_\mu^*) \cdot (\nabla \times \Hb),
\end{equation}
which can then be evaluated using eq. (\ref{eqn:EfromH}), and eqs. (\ref{eqn:ETE}) and (\ref{eqn:ETM}) for the various polarization combinations. The matrix elements are
\begin{align}
    \label{eqn:te-te}
    \mathcal{H}^\mathrm{TE-TE}_{\mu\nu} &= \oct{\omega_\mu^2} \oct{\omega_\nu^2} \eg \cdot \hat{e}_{\gb'} \times \sum_{j=0}^{N+1} \epsb_j^2 \eta_{j\Gb\Gb'} \times \big[A_{j\mu}^*A_{j\nu} I_j(-\chi_{j\mu}^* + \chi_{j\nu}) + B_{j\mu}^*B_{j\nu} I_j(\chi_{j\mu}^* - \chi_{j\nu}) \\ \nonumber & \hspace{4.6cm} + A_{j\mu}^*B_{j\nu} I_j(-\chi_{j\mu}^* - \chi_{j\nu}) + B_{j\mu}^*B_{j\nu} I_j(\chi_{j\mu}^* + \chi_{j\nu}) \big],
    \\ \label{eqn:tm-tm}
    \mathcal{H}^\mathrm{TM-TM}_{\mu\nu} &=
    \sum_{j=0}^{N+1} \eta_{j\Gb\Gb'} \times \big[\left(\gh \gh' \chi_{j\mu}^* \chi_{j\nu} + g g' \right) \left(A_{j\mu}^* A_{j\nu} I_j(-\chi_{j\nu} + \chi_{j\mu}^*) + B_{j\mu}^*B_{j\nu} I_j(\chi_{j\nu} - \chi_{j\mu}^*)\right) \\ \nonumber
    & \hspace{2cm} -\left(\gh \gh' \chi_{j\mu}^* \chi_{j\nu} - g g' \right) \left(A_{j\mu}^* B_{j\nu} I_j(-\chi_{j\nu} - \chi_{j\mu}^*) + B_{j\mu}^*A_{j\nu} I_j(\chi_{j\nu} + \chi_{j\mu}^*)\right) \big], 
    \\ \label{eqn:te-tm}
    \mathcal{H}^\mathrm{TE-TM}_{\mu\nu} &= i\oct{\omega_\mu^2} \eg \cdot \gh \times  \sum_{j=0}^{N+1} \epsb_j \eta_{j\Gb\Gb'} \chi_{j\nu} \times \big[-A_{j\mu}^*A_{j\nu} I_j(-\chi_{j\mu}^* + \chi_{j\nu}) + B_{j\mu}^*B_{j\nu} I_j(\chi_{j\mu}^* - \chi_{j\nu}) \\ \nonumber & \hspace{4.6cm} + A_{j\mu}^*B_{j\nu} I_j(-\chi_{j\mu}^* - \chi_{j\nu}) - B_{j\mu}^*B_{j\nu} I_j(\chi_{j\mu}^* + \chi_{j\nu}) \big],
\end{align}
and $\mathcal{H}^\mathrm{TM-TE}_{\mu\nu} =  (\mathcal{H}^\mathrm{TE-TM}_{\nu \mu})^*$. Note that eqs. (\ref{eqn:te-te}-\ref{eqn:te-tm}) are valid both for the overlap between two guided modes and for the overlap between a guided and a radiative mode.

\end{document}